# SymTC: A Symbiotic Transformer-CNN Net for Instance Segmentation of Lumbar Spine MRI

*Jiasong Chen [a], Linchen Qian [a], Linhai Ma [a], Timur Urakov [b], Weiyong Gu [c],* and *Liang Liang [a]*

[a] *Department of Computer Science, University of Miami, Coral Gables, FL*
[b] *Department of Neurological Surgery, University of Miami, Coral Gables, FL*
[c] *Department of Mechanical and Aerospace Engineering, University of Miami, Coral Gables, FL*

For correspondence:

Liang Liang, Ph.D.

Department of Computer Science

University of Miami

Ungar Building, Room 330K

Coral Gables, FL, 33146

Tel: (305) 284-8381; Email: liang@cs.miami.edu

# Abstract

Intervertebral disc disease, a prevalent ailment, frequently leads to intermittent or persistent low back pain, and diagnosing and assessing of this disease rely on accurate measurement of vertebral bone and intervertebral disc geometries from lumbar MR images. Deep neural network (DNN) models may assist clinicians with more efficient image segmentation of individual instances (disc and vertebrae) of the lumbar spine in an automated way, which is termed as instance image segmentation. In this work, we proposed SymTC, an innovative lumbar spine MR image segmentation model that combines the strengths of Transformer and Convolutional Neural Network (CNN). Specifically, we designed a parallel dual-path architecture to merge CNN layers and Transformer layers, and we integrated a novel position embedding into the self-attention module of Transformer, enhancing the utilization of positional information for more accurate segmentation. To further improve model performance, we introduced a new data augmentation technique to create synthetic yet realistic MR image dataset, named SSMSpine, which is made publicly available. We evaluated our SymTC and the other 15 existing image segmentation models on our private in-house dataset and public SSMSpine dataset, using two metrics, Dice Similarity Coefficient and 95% Hausdorff Distance. The results show that our SymTC has the best performance for segmenting vertebral bones and intervertebral discs in lumbar spine MR images. The SymTC code and SSMSpine dataset are publicly available at https://github.com/jiasongchen/SymTC.

**Keywords:** Lumbar spine MRI, Medical image instance segmentation, Transformer, Position embedding, Data augmentation

# 1. Introduction

The intervertebral discs in humans can undergo a profound degenerative process as early in the adolescence [1,2], which can be accompanied by facet arthropathy and hypertrophy. This degeneration can manifest as various conditions, including discogenic low back pain, disc herniation, spinal stenosis, and spondylolisthesis, which may necessitate the implementation of surgical or non-surgical interventions aimed at alleviating pain and restoring normal functionality. Magnetic resonance imaging (MRI) is the most widely used technique for specifically quantifying intervertebral discs degeneration (IDD) by assessing changes in disc geometry deformation and signal strength degradation [3–5]. The information derived from imaging data is of utmost importance for medical professionals in terms of both diagnosing medical conditions and planning appropriate treatments. Furthermore, this information serves as a critical foundation for developing patient-specific computational models, which hold the potential to mature over time and eventually enable accurate predictions of treatment outcomes within clinical settings. Presently, the process of geometry reconstruction, signal measurements, and grading from magnetic resonance (MR) images heavily relies on manual annotation. However, this process is not only time-consuming but also vulnerable to human bias. Consequently, there is an urgent need for automated MR image analysis methods to address these challenges.

In medical imaging, semantic/instance image segmentation, which divides the images into distinct sections at the pixel level so that each pixel belongs to a specific region, has the potential to be carried out through automated techniques [6].  The traditional methods, such as watershed and level set, have demonstrated satisfactory performance in medical image segmentation tasks. The watershed method treats an image as a topological map where intensity represents the altitude of the pixels. The watershed segmentation is determined by the watershed lines on a topographic surface [7,8]. The level set method performs image segmentation by utilizing dynamic variational boundaries [9]. However, the traditional method suffers from the clinical variation of different patients and the noise effect of different medical imaging equipment, resulting in issues such as the over-segmentation [10].

Since the increasingly vast amount of medical imaging data and computational resources have become available, machine learning (ML) methods, especially deep neural network techniques, show superior performance than traditional methods. Convolutional neural network (CNN) has a significant edge over its predecessors in that it possesses the capability to recognize essential components/features without requiring any human intervention [11]. CNNs are specifically designed to effectively utilize spatial and configural information by accepting 2D or 3D images as input. This approach helps to prevent the loss or disruption of structural and configural information in medical images [12]. Various deep CNNs, including UNet++ [13], Attention U-Net [14], MultiResUNet [15] and UNeXt [16] have been proposed for image segmentation for different medical imaging modalities and different organs [e.g., heart [17–19], lung [13], brain [20–23], pancreas [14], gland [23,24], spine [25,26], retina blood vessels [27,28], aorta [29–31], etc]. Although these methods have achieved promising performance, there are still some limitations in a more complex context coping with long-range dependency explicitly due to the intrinsic locality of convolutions.

Recently, Transformer, an ML technique, has shown exceptional performance not only on natural language processing (NLP) challenges like machine translation [32], but also image analysis tasks including image classification [33], image registration[34], image reconstruction[35], and image segmentation[24,36–42]. Various variations of Transformer models have demonstrated that the global information perceived by the self-attention operations is beneficial in medical imaging tasks. TransUNet was the first Transformer-based network specifically for medical image segmentation on the synapse multi-organ segmentation dataset [36]. Wang et al. [24] substituted the original skip connection scheme of U-Net with the proposed UCTransNet that includes a multi-scale Channel Cross fusion Transformer and a Channel-wise Cross-Attention and tested the network on the gland segmentation dataset [43] and synapse multi-organ segmentation dataset [44]. Hatamizadeh et al. [20,21] proposed both UNETR and Swin UNETR for 3D medical imaging segmentation. UNETR utilizes a U-shape network with a vision Transformer as the encoder and a CNN-based decoder. Swin UNETR is constructed by replacing the vision transformer encoder in UNETR architecture with the Swin Transformer encoder. Feng et al. [42] proposed SLT-

Net to utilize CSwin Transformer [46] as the encoder for feature extraction and the multi-scale context Transformer as the skip connection for skin lesion segmentation. Swin-Unet adopted Swin Transformer [37] with shifted windows as encoder and a symmetric Swin Transformer-based decoder with patch expanding layer as decoder for multi-organ segmentation task [17]. Pu et al. [44] proposed a semi-supervised learning framework with Inception-SwinUnet adopting convolution and sliding window attention in different channels for vessel segmentation on small amount of labeled data. Besides the self-attention mechanism, position embeddings are another crucial component of Transformer models. Regarding changing the order of the input, a Transformer model is invariant [32] without position embeddings. However, since text data inherently has a sequential structure, the absence of position information results in the ambiguous or undefined meaning of a sentence [48]. For image segmentation, usually, an image patch is treated as a token, and Transformers process the entire input sequence of tokens in parallel. With position embeddings, a Transformer would be able to differentiate between image patches with similar content that appear in different positions in the input image, which is beneficial for image segmentation applications. A variety of different methods may be used to incorporate the position information into Transformer models. Absolute position encoding and relative position encoding are two main categories to encode a token's position information. Vaswani firstly introduced absolute and relative position embedding in the vanilla Transformer model [32]. Shaw extended the self-attention mechanism with the capacity of effectively incorporating the representation of relative position [49]. Valanarasu et al. [23] proposed a gated position-sensitive axial attention mechanism to cope the difficulty in learning position encoding for the images.

Specifically for lumbar spine research, instance segmentation of MR images is preferred, which not only determines whether or not a pixel belongs to a disc, but also labeling the precise instance to which it belongs [6]. In recent years, most instance segmentation methods for spine image segmentation are based on CNNs-only networks, and only a few Transformer-based networks are employed. For example, Kuang et al. [47] built an unsupervised segmentation network for spine image segmentation using the rule-based region of interest (ROI) detection, a voting mechanism accompanied by a CNN network. Sekuboyina et al. [25] proposed a dual branch fully convolutional network that takes advantages of both low-resolution attention information on two-dimensional sagittal slices and high-resolution segmentation context on three-dimensional patches for effective segmentation of the vertebrae. MLKCA-Unet incorporates multi-scale large-kernel convolution and convolutional block attention into the U-net architecture for efficient feature extraction in spine MRI segmentation [26]. Pang et al. [48] introduced a mixed-supervised segmentation network and it was trained on a strongly supervised dataset with full segmentation labels and a weakly-supervised dataset with only key points. BianqueNet combined new modules with a modified deeplabv3+ network [52], which includes a Swin Transformer-skip connection module, for segmentation of lumbar intervertebral disc degeneration related regions [53].

It was shown that the Transformer-based models only perform effectively when trained on large-scale datasets since the lack of inductive bias [54]. The utilization of Transformer-based networks for medical imaging tasks poses a challenge due to the limited availability of labeled images in medical datasets. Obtaining well-annotated medical imaging datasets presents significantly greater challenges compared to curating traditional computer vision datasets. Dealing with expensive imaging equipment, complex image acquisition pipelines, expert annotation requirements, and privacy concerns are all part of the problematic issues [55]. This scarcity hampers the effective application of Transformer-based models in the medical domain. In such scenarios, the adoption of suitable and feasible data augmentation techniques becomes crucial prior to model training. These techniques can help to increase the effective size of the medical image dataset and improve the performance of the Transformer-based model.

In this study, we propose Symbiotic Transformer-CNN (SymTC) network, an innovative model that effectively combines CNN and Transformers for the automatic lumbar spinal MRI instance segmentation, together with a novel data synthesis method based on statistical shape model (SSM) and biomechanics. For SymTC, a novel relative position embedding (RPE) is proposed for segmentation performance improvement. The SSM-biomechanics-based data synthesis method could generate lumbar spine images with large and plausible deformations, which can be used for model training and evaluation. The code and the generated lumbar spine dataset (SSMSpine) will be publicly available at GitHub when the paper is published.

## 2. Related Work

*2.1. Transformer-based Networks for the segmentation of non-spine medical images*

Transformer-based networks have shown promising performance on medical image segmentation because of their ability to capture long-range dependencies. Comparably, the inductive bias of CNN networks benefits from its local connectivity and parameter sharing property. Therefore, many networks, combining CNN and Transformers and leveraging both benefits, have been proposed in the past few years. TransUNet [36] firstly combines CNNs and Transformer in a cascaded manner in its encoder for medical imaging tasks, in which the low-level features are collected from the CNNs and then fed to the Transformer to capture global interactions. Other designs of CNN and Transformer networks, including UNETR [21], UTNet [18], UCTransNet [24], Swin UNETR [20] and MedT [23] have shown better segmentation performances in different medical image modalities, compared to CNN-only networks. UNETR replaces the encoder of a UNet with Transformer layers with 1D learnable positional embedding, and its decoder only has convolution layers [21]. UTNet inserts Transformer layers into a Unet in a sequential manner: a convolution layer followed by a Transformer layer with learnable relative position embedding [18]. UCTransNet embeds Transformer layers into the skip-connections of a Unet with fully learnable absolute position embedding [24]. Swin UNETR replaces the Transformer in UNETR with Swin-Transformer [20]. Medical Transformer (MedT) uses gated-axial Transformer layers in the encoder of a Unet [23]. HSNet used PVTv2 [56] as encoder and a dual-branch structure which Transformer branch and CNN branch fused by element-wise product as decoder for polyp segmentation [57].

*2.2. Transformer-based Networks for the segmentation of spine images*

To the best of our knowledge, there are only a few Transformer-based networks [58,59] specifically designed for lumbar spine image segmentation, including EG-Trans3DUNet [58], Spine-transformers [59], APSegmenter [60], and BianqueNet [53]. However, it is worth noting that most of these networks were developed for the segmentation of vertebral bodies using CT modality, which may not be directly applicable to the segmentation of the intervertebral discs in order to study disc degeneration. Among those networks, only the code of BianqueNet is publicly available.

EG-Trans3DUNet combines two vision transformer branches to handle both local patches and resized global spinal CT images [58], and it merges edge characteristics and semantic features generated by a CNN-based edge detection block. Spine-transformers were designed for spinal CT image segmentation with a two-stage pipeline to handle arbitrary Field-Of-View input images [59]. In its first stage, a Transformer with a CNN backbone is utilized as a 3D object detector to locate individual vertebrae, and then the input image is cropped into regions of individual vertebrae. In its second stage, a multi-task encoder-decoder CNN network is applied to each cropped region to segment the vertebra. The source code of the second stage is not publicly available. APSegmenter [60] combines a ViT-style Transformer with a mask Transformer to segment spine X-ray images, and an adaptive postprocessing is applied to further refine the result. BianqueNet [53] employed a resnet101 network to perform feature extraction, followed by upsampling using the Swin Transformer-skip connection module and a double upsampling operations. It also used a multi-scale feature fusion module to generate the segmentation of regions associated with intervertebral disc degeneration.

In addition to the segmentation of vertebral bodies, the segmentation of intervertebral discs (IVDs) is vital for lumbar spinal disease diagnosis and treatment. Since the water content in IVDs cannot be revealed on CT images, currently, MRI is the gold standard imaging modality for the evaluation of IVD pathologies [61]. Our study aims to explore the benefit of combining CNN and Transformer for instance segmentation of lumbar spine MR images.

## 2.3. Self-Attention in Transformer

As opposed to convolutional operations, the self-attention mechanism within a Transformer network has the fundamental advantage of effectively capturing global features and long-range contextual dependency. It uses the Key, Query and Value vectors to better describe the features' connections. Nonetheless, because of the inherited properties of self-attention, it does not retrieve the position information on its own, which is important for instance segmentation. One of the best ways to tackle this problem is to use a well-designed position embedding mechanism to inject the position relationships into the self-attention calculation.

### 2.3.1. The classic self-attention mechanism with additive position embedding

The plain Transformer is constructed with the multi-head self-attention modules (MHSA), which enable Transformer to capture and utilize more accurate and detailed spatial information [32]. Given an input token set (e.g., image patches) $X$, three individual linear transformations ($W_Q, W_K, W_V$) are applied to $X$ to generate query embedding ($Q$), key embedding ($K$), and value embedding ($V$). Then, the self-attention score, $Attn$, is calculated as a scale-product of these three embedding as following:

$$Attn = softmax\left(\frac{QK^T}{\sqrt{d}}\right) \quad (1)$$

$$Out = Attn \times V \quad (2)$$

In the above equations, $Q = (X + P)W_Q$, $K = (X + P)W_K$, $V = (X + P)W_V$. $P$ is the encoded position, and $d$ is the dimension of embedding in each head. $Out$ is the final output of the self-attention module.

### 2.3.2. Position Embedding

Without using any position embedding, the self-attention mechanism in Eq.(1) is permutation-invariant and cannot distinguish tokens (e.g., image patches) at different spatial locations. Therefore, it is essential to design efficient position embedding cooperating with the self-attention mechanism.

The attention matrix in the classic self-attention Eq.(1) can be decomposed into three terms:

$$Attn \sim QK^T = XW_Q(XW_K)^T + P\,W_Q(XW_K)^T + XW_Q(PW_K)^T + PW_Q(PW_K)^T \quad (3)$$

                Content-Content     Content-Position     Position-Position

Therefore, the attention considers three interactions among tokens: $XW_Q(XW_K)^T$ for content to content interaction, $P\,W_Q(XW_K)^T + XW_Q(PW_K)^T$ for interaction between content and position, and $PW_Q(PW_K)^T$ for position to position interaction. As the instance segmentation task is location-specific, a well-designed interaction between content and position could improve self-attention ability to utilize both content and position information to accomplish the instance segmentation task.

Generally, there are mainly two steps to define the position embedding. The first step is defining the position function or distance function, which is used for encoding the position information of input tokens. There are plenty of position functions such as index function, Euclidean distance, and sinusoidal functions etc., The second step is defining methods to incorporate the encoded position information into self-attention.

Absolute position embedding and relative position embedding are two main position representation methods to incorporate the position information into input tokens. Absolute position embedding encodes the absolute positions of each input tokens as individual encoding vectors, and relative position embedding focuses on the relative positional relationships of pairwise input tokens [62,63]. In the vanilla Transformer designed for NLP [32], it used a combination of absolute and relative position embedding to add position information to the tokens. It is inconclusive whether relative position embedding is better or worse than absolute position embedding, and the answer seems to be dependent on specific applications [48,49,63,64]. Relative position encoding benefits from capturing the details of relative distance/direction and is invariant to tokens' shifting. The intuition is that, in the self-attention mechanism, the pairwise positional relationship (both in terms of direction and distance) between input elements might be more advantageous than absolute position of individual elements [62]. In such a case, position information in Transformer is an extensive research area, and various relative position encodings have been proposed for medical imaging segmentation [54]. For example, UTNet proposed the 2-dimensional relative position encoding by adding relative height and width information [18]. MedT updated self-attention mechanism with position encoding along the width axis with the inspiration of axial attention [23,65,66]. The Parameter-Efficient Transformer added a trainable position vector to the input to encode relative distances [22]. In this work, we propose a novel relative position embedding method for segmentation performance improvement.

### 2.3.3. Self-Attention in the Existing Image Segmentation Models

We compared our model with 15 representative image segmentation models on our dataset, including 11 Transformer-based models and 4 CNN-only models. Our model is significantly different from the others, not only in network structure but also in the self-attention mechanism.

Table 1 summarizes the self-attention mechanisms employed in the existing 11 Transformer-based models. TransUnet and UNETR add the position embedding directly into the input patches [21,36]. Swin-Unet, Swin UNETR, and Inception-SwinUnet incorporate the position embedding into attention score instead of input tokens. [17,20,47]. BianqueNet utilized the position embedding within Swin-Transformer [53]. According to the source code, SLT-Net introduces a lepe distance to represent the position bias (embedding), which is directly added into the output matrix [45]. MedT proposed a gated position-sensitive axial attention mechanism where four learnable gates ($G$) control the amount of position embedding contained in key ($K$), query ($Q$) and value ($V$) embeddings [23]. UTNet introduced the 2-dimensional relative position encoding by adding relative position logits along height and width dimensions($R_{height}$, $R_{width}$) into the key embedding [18]. HSNet and UCTransNet do not include position embedding in their models [57,24].

**Table 1.** Self-Attention in the existing attention/Transformer-based segmentation models

| Model | Self-Attention |
|---|---|
| Swin UNETR | $softmax\left(\frac{XW_Q(XW_K)^T}{\sqrt{d}} + R\right)(XW_V)$ <br> Note: $R$ is relative position bias in the reference |
| SLT-Net | $softmax\left(\frac{XW_Q(XW_K)^T}{\sqrt{d}}\right)(XW_V) + lepe(XW_V)$ <br> Note: $lepe$ is a convolution kernel |
| UNETR | $softmax\left(\frac{(X+P)W_Q((X+P)W_K)^T}{\sqrt{d}}\right)(X+P)W_V$ |
| Inception-SwinUnet | $softmax\left(\frac{XW_Q(XW_K)^T}{\sqrt{d}} + R\right)(XW_V)$ <br> Note: $R$ is relative position bias in the reference |
| HSNet | $softmax\left(\dfrac{XW_Q(XW_K)^T}{\sqrt{d}}\right)(XW_V)$ |
| Swin-Unet | $softmax\left(\dfrac{XW_Q(XW_K)^T}{\sqrt{d}} + R\right)(XW_V)$<br>Note: $R$ is relative position bias in the reference |
| TransUNet | $softmax\left(\dfrac{(X+P)W_Q((X+P)W_K)^T}{\sqrt{d}}\right)(X+P)W_V$ |
| MedT | $softmax\left(XW_Q(XW_K)^T + G_Q XW_Q(R_Q)^T + G_K XW_K(R_K)^T\right)\left(G_{V_1} XW_V + G_{V_2} R_V\right)$<br>Note: $G_Q, G_K, G_{V_1}, G_{V_2}$ are learnable parameters for gating mechanism. $R_Q, R_K, R_V$ are the position bias for Query, Key, and Value |
| UTNet | $softmax\left(\dfrac{XW_Q(XW_K + R_{width} + R_{height})^T}{\sqrt{d}}\right)XW_V$<br>Note: $R_{width}$ and $R_{height}$ are the relative height and width information |
| UCTransNet | $softmax\left(\dfrac{X_c W_Q(X_c W_K)^T}{\sqrt{d}}\right)(X_c W_V)$<br>Note: $X_c$ is composed of image channels instead of image patches. |
| BianqueNet | $softmax\left(\dfrac{XW_Q(XW_K)^T}{\sqrt{d}} + R\right)(XW_V)$<br>Note: $R$ is relative position bias in the reference |

## 3. Methods

In this section, we depict the overall architecture of our SymTC network, and explain the pipelines and modules in detail, including TC modules, merge modules, novel self-attention with relative position embedding, and data augmentation. At the end, we provide the loss function for model training.

### 3.1. The architecture of our SymTC network

The overall architecture of our network is shown in Figure 1, including TC modules (TCM), Merge modules, and a segmentation head. TC modules utilize Transformer and CNN synthetically, which serve as a feature extractor for yielding hierarchies of features. The Merge modules fuse multi-scale feature maps. The segmentation head is used for generating the final instance segmentation masks. The overall diagram of SymTC resembles a U-Net [67], and novelties are in the TC modules that set out network apart from the others.

The network works as follows. First, a 2D sagittal slice of spine MR image $I \in \mathbb{R}^{1 \times 512 \times 512}$ is processed along the encoder path. From the encoder part, we can obtain three levels of semantic features $A_i \in \mathbb{R}^{C_i \times \frac{H}{S_i} \times \frac{W}{S_i}}$, where $i \in \{1, 2, 3\}$, $C_i \in \{32, 128, 512\}$ and $S_i \in \{1, 4, 16\}$. Subsequently, the bottleneck feature is sent to the decoder path for progressively upsampling. The lower-level hierarchical features $A_1$, $A_2$, and $A_3$ and the corresponding high-level features $B_1$, $B_2$ and $B_3$ at the decoder side are then fused in the Merge modules for obtaining the pyramid feature $M_i \in \mathbb{R}^{C_i \times \frac{H}{S_i} \times \frac{W}{S_i}}$. Thereafter, the final feature $M_1$ is sent into a refine module (TC module) followed by the segmentation head to obtain the prediction masks $\hat{Y} \in \mathbb{R}^{12 \times 512 \times 512}$.

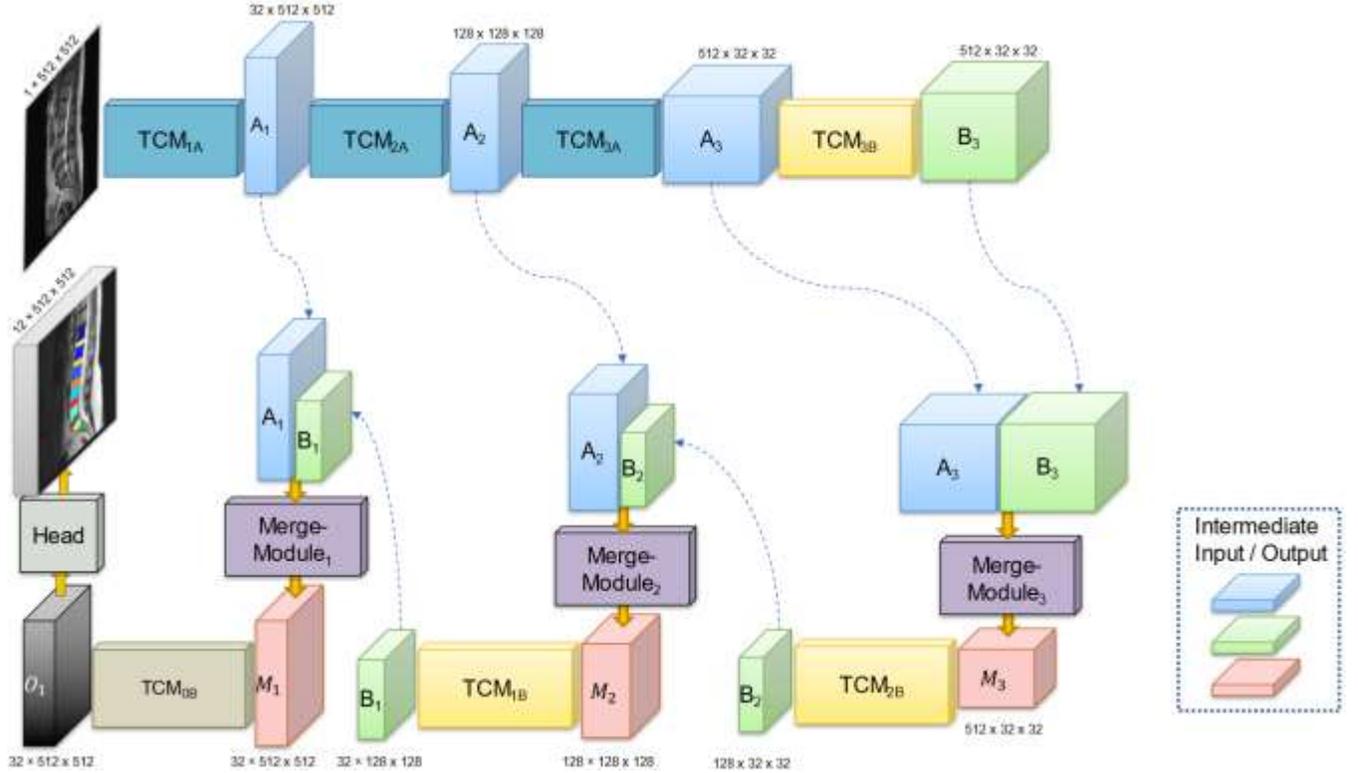

**Fig. 1.** The architecture of SymTC

*3.2. TC Module with parallel CNN and Transformer paths*

Based on the two properties of CNN and Transformer: (1) CNN can model the interaction between adjacent features in the same receptive field via a shifted convolution window. (2) Transformer has the excellent ability for incorporating global features and perceiving long-range dependencies, some networks have been proposed for fusing CNN and Transformer to combine their advantages. Inspired by TransUnet for combining CNN and Transformer in a serial order, we introduce a symbiotic module (TC module) for combining two parallel CNN and Transformer paths. The proposed TC Module integrates the advantages of both CNNs and Transformers to enhance the capability not only to capture more accurate and detailed structural and spatial information, but also to distinguish the boundaries between vertebral bodies and intervertebral discs (IVDs). The architecture of TC module is shown in Figure 2. There are two paths taking the same input in a TC module, including CNN path and Transformer path.

The CNN path in the TC module is a residual block consisting of 5x5, 1x1 and 3x3 convolutional filters along with shortcut/skip connections, which is the basic ResNet architecture. The Transformer path in TC module follows the basic architecture of ViT [54]. The input goes to convolutional patch embedding and then is sent to $N$ layers of Transformer blocks, each block adopts a modified multi-head self-attention layer with relative position embedding (RMHA) for feature extraction and a feed-forward layer. Since we give equal importance to CNN and

Transformer, the two parallel paths work synthetically and mutualistically. Subsequently, feature maps of two paths are concatenated and fed to norm and ReLU layers.

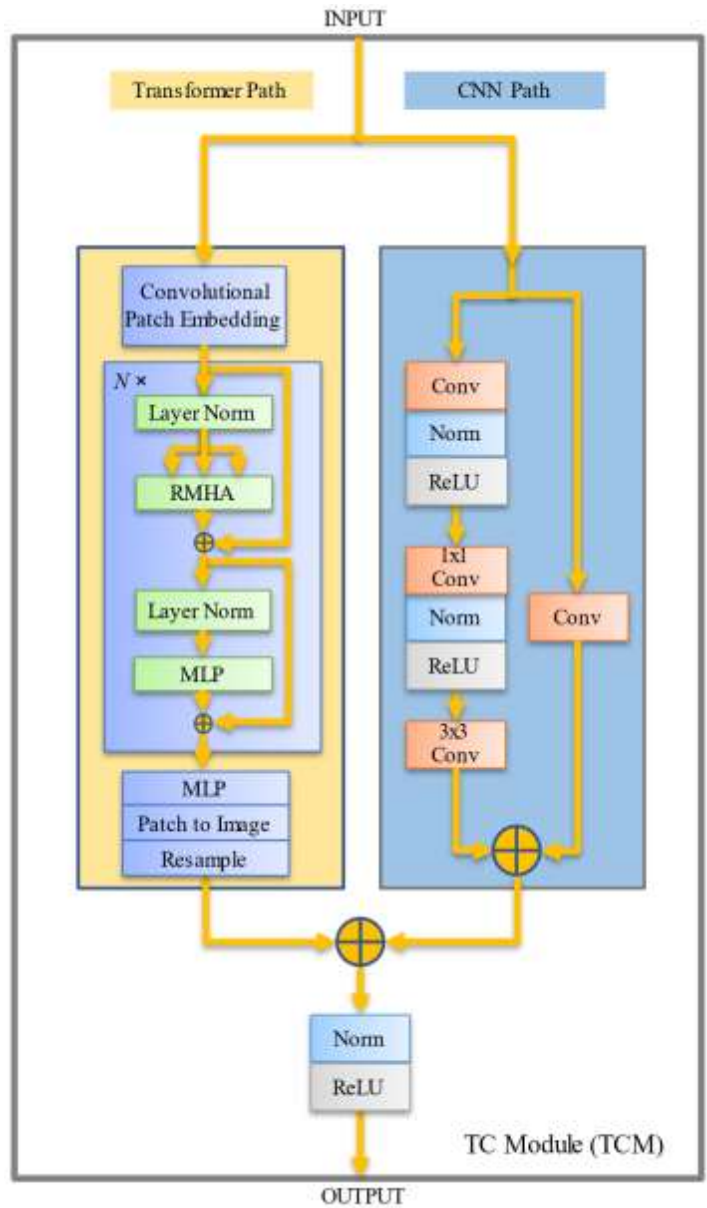

**Fig. 2**. The architecture of TC Module in SymTC, which consists of a CNN path (upper) for local features and a Transformer path (lower) for generating global features. "Norm" refers to GroupNorm or LayerNorm.

Our default configuration of Transformer path in the TC module consists of 2 Transformer layers (L=2), an embedding size of 512, and 16 attention heads. We employed three different patch sizes for the TC module: 16×16, 4×4 and 1×1. Specifically, the resolution of 16x16 was applied to the first and second TC module, while the resolutions of 4×4 and 1×1 were utilized in the third and fourth blocks, respectively. This variation in patch sizes allows for capturing different levels of details and context within the input data across different stages of the network.

*3.3. Novel Self-Attention with Relative Position Embedding in SymTC*

We propose a novel relative position embedding to improve segmentation performance. We redefine the attention score matrix by modifying the content-content and content-position components, while eliminating the position-position component, as delineated in Eq.(4):

$$A = softmax\left(\frac{XW_Q(XW_K)^T R_1 + XW_Q R_2}{\sqrt{d}}\right) \quad (4)$$

In the above Eq.(4), $R_1$ and $R_2$ are two matrices encoding relative positions between tokens, which are significantly different from the position encoding methods listed in Table 1. $X$ is a matrix containing all of the input tokens. The utilization of these relative position components may more effectively include position information. Consequently, the modified attention restructures the content-content and content-position components through the integration of our novel relative position embedding.

To better explain Eq.(4) with more details, we expand it from the perspective of individual tokens. For relative position encoding, we redefine query and key by incorporating the sinusoidal functions to effectively capture the relative position information, which is illustrated by Eq.(5) and Eq.(6).

$$q_i = \begin{bmatrix} x_i W_Q \odot \cos(p_i W_1 + b_1) \\ x_i W_Q \odot \sin(p_i W_1 + b_1) \\ x_i W_Q \odot \cos(p_i W_2 + b_2) \\ x_i W_Q \odot \sin(p_i W_2 + b_2) \end{bmatrix} \quad (5)$$

$$k_j = \begin{bmatrix} x_j W_K \odot \cos(p_j W_1) \\ x_j W_K \odot \sin(p_j W_1) \\ \cos(p_j W_2) \\ \sin(p_j W_2) \end{bmatrix} \quad (6)$$

In the above equations, $x_i$ and $x_j$ denote the input token-$i$ and the input token-$j$ (e.g., two image patches) in the format of row vectors. $p_i$ and $p_j$ are the absolute position vectors of token-$i$ and token-$j$. $W_1$ and $W_2$ are trainable weight matrices, and $b_1$ and $b_2$ are trainable bias vectors, which are used inside the sine and cosine functions to generate different frequency components. $\odot$ denotes element-wise multiplication between two vectors.

Each entry, $a_{ij}$, of the attention score matrix, $A$, is proportional to the scalar dot product of the query vector $q_i$ and the key vector $k_j$, which can be derived from the following three equations:

$$q_i \odot k_j = x_i W_Q \odot x_j W_K \odot \cos(p_i W_1 + b_1) \odot \cos(p_j W_1) + x_i W_Q \odot x_j W_K \odot \sin(p_i W_1 + b_1) \odot \sin(p_j W_1) \\ + x_i W_Q \odot \cos(p_i W_2 + b_2) \odot \cos(p_j W_2) + x_i W_Q \odot \sin(p_i W_2 + b_2) \odot \sin(p_j W_2) \quad (7)$$

$$dot(q_i, k_j) = sum(q_i \odot k_j)$$
$$= sum\left(x_i W_Q \odot x_j W_K \odot \cos\left((p_i - p_j)W_1 + b_1\right) + x_i W_Q \odot \cos\left((p_i - p_j)W_2 + b_2\right)\right) \quad (8)$$

$$a_{ij} = \frac{\exp[dot(q_i, k_j)]}{\sum_{j'} \exp[dot(q_i, k_{j'})]} \quad (9)$$

In the above equations, $sum$ computes the summation of all elements. The attention score between the two tokens, $a_{ij}$, is a function of the relative position vector $p_i - p_j$, which is modulated by cosine functions with different frequencies and phases. We add bias into the Eq.(5), so that the cosine functions in Eq.(8) may have different phases. The compound relative position encoding, which incorporates both sine and cosine functions in Eq.(5) and Eq.(6), effectively captures the relative position information of the input.

The final output of the attention module contains not only the original term in Eq.(2) but also a new term related to the relative positions. The *i*-th row of the final output is given by:

$$out_i = Linear((A_i X W_V)^T) + Linear\left(\sum_j a_{ij}(p_i - p_j)\right) \quad (10)$$

In the above Eq.(10), $A_i$ is the row-*i* of the attention score matrix $A$, $Linear$ denotes a linear layer with trainable weight and bias. We note that, in the actual implementation, $W_Q$ and $W_K$ are replaced by MLPs.

### 3.4. Novel data augmentation/synthesis method based on SSM and biomechanics

For medical image data augmentation, elastic deformation is often used for nonlinear deformation of the images to increase diversity of training data [67]. Briefly, the input space is discretized by a grid, and a random displacement field on the grid is generated by sampling from a normal distribution with standard deviation equal to $\sigma \times$ grid resolution (i.e., the size of a grid cell). The parameter $\sigma$ determines deformation magnitude. To ensure a large deformation with diffeomorphism, the grid needs to be coarser than the input size (i.e., 512 ×512). In this study, we applied two successive elastic deformations to each training image, with grid sizes of $9 \times 9$ and $17 \times 17$. As shown in Figure 3, when the deformation parameter $\sigma$ is larger than 0.5, the generated images and spine shapes are highly unrealistic. Figure 4 shows two impractical synthesis instances using elastic deformation with $\sigma$ is 0.5. To avoid unrealistic shapes, we used $\sigma = 0.25$ in the experiments.

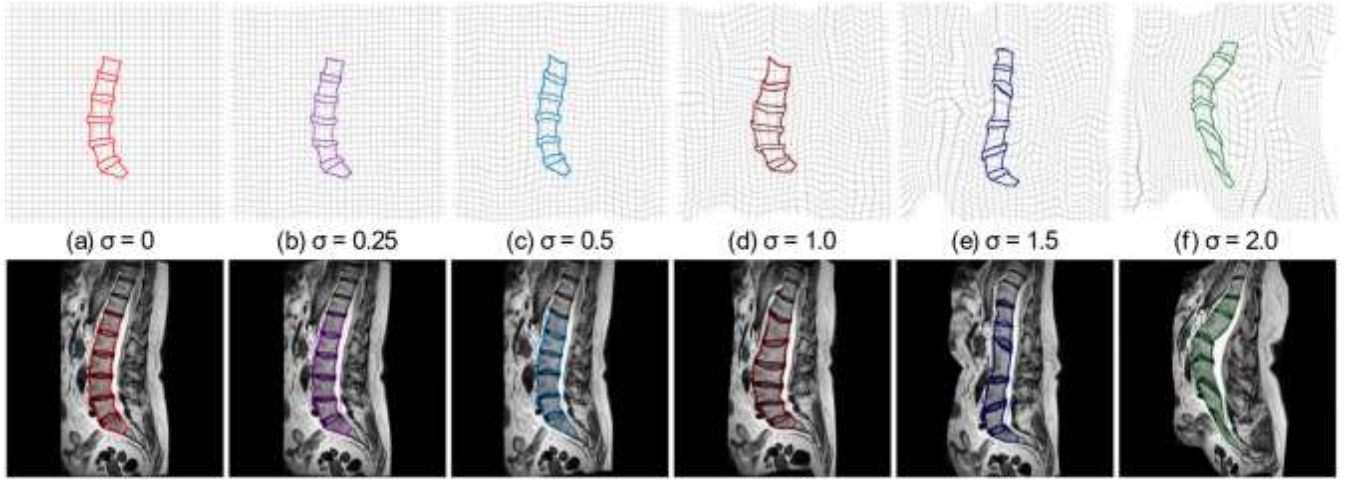

**Fig. 3.** Data augmentation/synthesis examples using elastic deformation with $\sigma$ from 0 to 2.0.

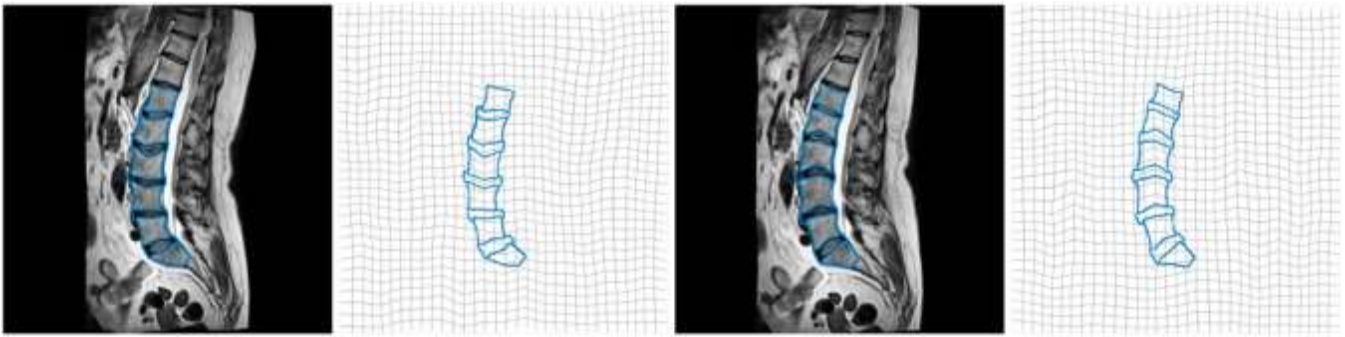

**Fig. 4.** Impractical synthesis instances using elastic deformation with $\sigma = 0.5$.

In this work, we developed a new method to synthesize lumbar spine MR images suitable for model training and evaluation. First, we built a statistical shape model (SSM) of lumbar spine shapes (i.e., contours of discs and vertebrae) in a dataset set, and the SSM represents the probability distribution of lumbar spine shapes. We refer the reader to the reference papers [68–72] for the details of constructing an SSM. By sampling from the SSM, different lumbar spine shapes can be generated, and each generated lumbar spine shape could be considered from a virtual patient. We note that the SSM technique has been used to generate virtual but realistic patient geometries in many applications, such as generating aortic aneurysm geometries [73–75]. Given a lumbar spine shape, if a lumbar spine MR image can be generated and consistent with the shape, then we will have a new sample with ground-truth. For this purpose, we developed a biomechanics-based method to generate a lumbar spine MR image $\tilde{I}$ from a lumbar spine virtual shape $\tilde{S}$ by using a reference image $I$ with its ground-truth shape $S$. Intuitively speaking, a nonlinear spatial transform from the shape $S$ to the virtual shape $\tilde{S}$ is determined by using biomechanics principles [76–78], and then $\tilde{I}$ is obtained by applying the spatial transform to $I$. The generated images are visually plausible, as shown in Figure 5.

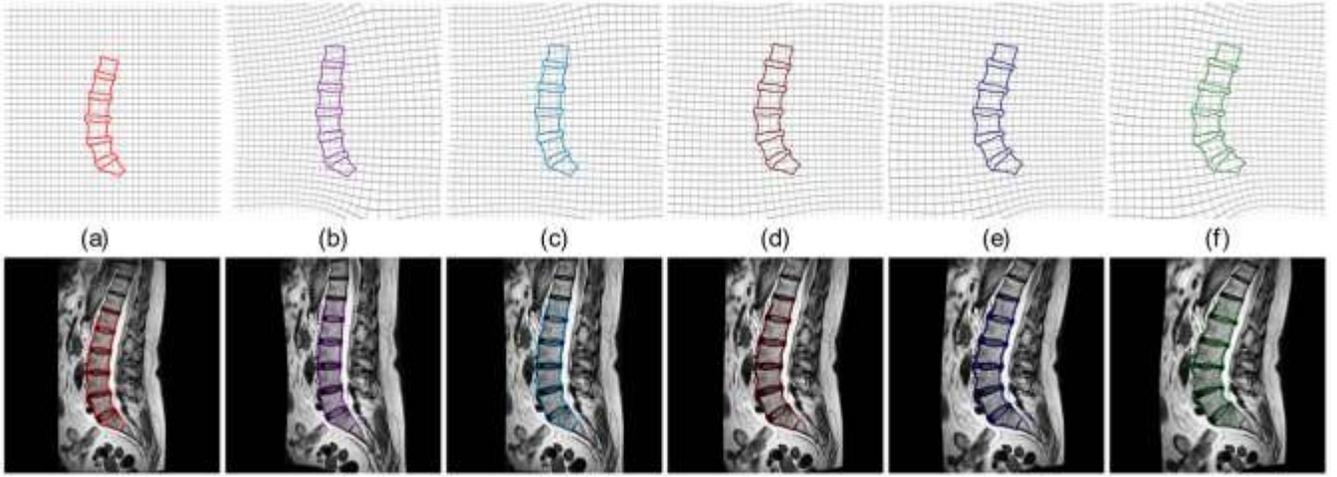

**Fig. 5.** data augmentation/synthesis examples (a-f) using our method. Please zoom in for better visualization.

In the implementation, we obtain the spatial transform $T$ from $\tilde{S}$ to $S$, and apply the spatial transform $T$ to a regular mesh grid around the virtual shape $\tilde{S}$ to obtain a deformed grid in the space of the reference image $I$, and then $\tilde{I}$ is obtained by interpolating pixel values of $I$ at each node of the deformed grid, i.e., $\tilde{I}(x,y) = I(T(x,y))$ where $(x,y)$ denotes a 2D spatial point and $T(x,y)$ is the transformed point. By using biomechanics and finite element analysis (FEA), the spatial transform, i.e., the deformation field on the mesh grid, is obtained by minimizing the following energy/loss function $\Pi$:

$$\Pi = \int_V \Psi \, dV + \lambda \cdot avg_i \|T(\tilde{S}(i)) - S(i)\|^4 \tag{11}$$

In the above Eq.(11), $V$ represents the undeformed mesh grid of the image $\tilde{I}$ to be generated, $\tilde{S}(i)$ is the $i$-th point location of the shape $\tilde{S}$, $T(\tilde{S}(i))$ is the transformed point location that needs to be equal to $S(i)$, and $\|\blacksquare\|$ denotes vector L2 norm. $avg$ is the average operator. $\lambda$ is a weight constant (set to 16 in experiments). $\Psi$ is the strain energy density function that is determined by deformation and mechanical property of soft biological tissues around the lumbar spine. From the perspective of FEA and biomechanics, the Eq.(11) simulates the scenario that under the external "force" proportional to $T(\tilde{S}(i)) - S(i)$ at each point of the lumbar spine shape, the soft biological tissues of human body will deform and reach to an equilibrium state of minimum energy. To speed up the optimization process, we use a deep neural network with sine activation functions to parameterize the transform $T$, i.e., $T(x,y) = DNN(x,y)$, and then the energy function in Eq.(11) becomes a function of the DNN internal parameters. The energy optimization problem is resolved by adjusting/optimizing the parameters of the DNN. Once the optimization is done, the deformation field is obtained and then the image $\tilde{I}$ is generated. Since our goal is to generate plausible images for model training and evaluation in the image segmentation tasks, not for patient-specific FEA simulation of human body deformation, we made an assumption about the strain energy density function to reduce computation cost: tissue mechanical behavior follows the Ogden hyperelastic model with homogeneous tissue properties [76,77]. The whole procedure is implemented by using our newly developed PyTorch-FEA library for large deformation biomechanics [78].

We observed that generative models, such as GAN [79,80] and diffusion [81,82], could be used to generate synthetic images for data augmentation purposes. However, the synthetic images generated by these models typically require manual annotation to obtain ground-truth labels, which is both time-consuming and labor-intensive. Our method not only generates realistic images but also produces the corresponding ground-truth labels (i.e., shapes).

*3.5. Loss function*

For each model, we use the original loss function if it is applicable to our application. If the original loss function is not suitable (e.g., it is for binary classification only), then we use the loss function $\mathcal{L}$ that combines a Dice loss $\mathcal{L}_{Dice}$ and an area-weighted cross entropy loss $\mathcal{L}_{aw\_ce}$.

$$\mathcal{L} = 0.5 L_{Dice} + 0.5 L_{aw\_ce} \quad (12)$$

$$\mathcal{L}_{Dice} = 1 - \frac{2\sum_{i,j}(y(i,j)\hat{p}_m(i,j)) + \epsilon}{\sum_{i,j}(y(i,j) + \hat{p}_m(i,j)) + \epsilon} \quad (13)$$

$$\mathcal{L}_{aw\_ce} = -\sum_{i,j,m} w_m y_m(i,j) \log(\hat{p}_m(i,j)) \quad (14)$$

$\hat{p}_m(i,j)$ is the m-th element in the output tensor from the softmax layer at the pixel location $(i,j)$, which corresponds to the m-th object (a disc or a vertebra) at the location $(i,j)$. $y(i,j)$ is the true label of the pixel at location $(i,j)$. $w_m$ is a nonnegative weight inversely proportional to the area of the m-th object, and $\sum_m w_m = 1$. $\epsilon$ is a small constant (1e-4) to prevent the case of 0/0 in Eq.(13). In a lumbar spine image, where the background area is substantially larger than the combined area of the discs and vertebrae, employing area-weighted cross entropy loss effectively reduces the influence of the background in the loss function.

## 4. Experiments

*4.1. Original Dataset and Augmented Datasets*

Our dataset consists of a total of 100 patients' lumbar spine T2-weighted MR images from the University of Miami medical school, with personal identification information removed. Following the protocol in [83], five lumbar discs (D1, D2, D3, D4, D5) and six vertebral bones (L1, L2, L3, L4, L5, S1) in each patient's mid-sagittal MR image was manually annotated by three trained operators to identify and mark the boundaries and landmarks of the lumbar discs and vertebrae. To ensure accuracy and consistency, the three operators engaged in discussion to reach a consensus on the best annotation (i.e., ground-truth), for each mid-sagittal MR image. In the literature, lumbar disc D1 is also called L1/L2, similarly D2 for L2/L3, D3 for L3/L4, D4 for L4/L5, and D5 for L5/S1. The MR images are of various resolutions, and each of the images are resized to 512 × 512. Each image is also pre-processed independently by normalizing the intensities into range [0,1]. The average pixel spacing is 0.7004 mm. The dataset of 100 patients was divided into 70 training samples, 10 validation samples, and 20 test samples, and those samples are referred to as the original training/validation/test samples in this paper.

Subsequently, the augmented dataset, named SSMSpine, is generated by using our method in Section 3.4. The SSMSpine dataset not only contains realistic lumbar spine MR images but also includes the corresponding ground-truth labels. The SSMSpine dataset is divided into three sets: an augmented training set with 7000 samples, an augmented validation set with 250 samples, and an augmented test set with 2500 samples. To generate the augmented test set, an SSM was constructed using the 20 original test samples, and then 125 virtual shapes were generated from the SSM. Using each of the 20 original test samples as a reference image and each of the 125 virtual shapes, $20 \times 125$ (=2500) new MR images were generated using the method in Section 3.4. The augmented training and validation sets were generated in a similar way using another SSM built on both the original training and validation samples. 70 original training samples and 100 virtual shapes were used to generate the augmented training set, consisting of 7000 samples. 10 original validation samples and 25 virtual shapes were used to generate the augmented validation set, consisting of 250 samples. More information about the SSMSpine dataset is available on GitHub.

*4.2. Model evaluation and comparison*

In our study, we trained and compared a total of 16 models on our lumbar spine MRI dataset. These models are SymTC (ours), Attention U-Net [14], HSNet [57], Inception-SwinUnet [47], MedT [23], MultiResUNet [15], SLT-Net [45], Swin-Unet [17], UNETR [21], Swin UNETR [20], TransUNet [36], UCTransNet [24], UNet++ [13], UNeXt [16], UTNet [18], and BianqueNet [53]. Due to either unavailability (e.g., no source code) or incompatibility (e.g., size not matching), we were unable to test all models mentioned in Section 2 **(Related Work)**. By training and evaluating these models, we aimed to compare their performance and determine the most effective approach for spine MRI instance segmentation on our dataset. To ensure compatibility with our dataset, we made minor adjustments to the original codes of some models if necessary.

We conducted two experiments: experiment-A and experiment-B. In experiment-A, each model is trained using the original training set with elastic deformation (Section 3.4). The top 5 models in experiment-A are selected for training using the augmented training set in experiment-B. In both experiments, we applied random translations to the input images within 16 pixels during training, which is a common data augmentation method. In both experiments, the augmented validation set is used for hyper-parameter tuning. The evaluation process is divided into two parts: instance segmentation evaluation and translation robustness evaluation.

In the instance segmentation evaluation, we assess model segmentation performance for individual lumbar spine instances/objects in the MR images. The instance segmentation task is formulated as a task of labeling 12 distinct objects (5 lumbar discs, 6 vertebrae, and 1 background). The input to each model is a single-channel mid-sagittal lumbar spine MR image with size of $512 \times 512$ pixels. In the segmentation output, each class is represented by a distinct channel as a binary segmentation map. We employ both the Dice Similarity Coefficient (DSC) and the 95% Hausdorff Distance (HD95) as evaluation metrics. In both experiments, the original test set with 20 samples and the augmented test set with 2500 samples are used separately for model performance assessment on unseen data.

Apart from assessing the instance segmentation performance, we also assess model robustness against translation transformation to the images, i.e., translation robustness. We quantify a model's translation robustness by evaluating its segmentation performance on the test set subjected to different translations in the vertical direction and horizontal direction. This part of the evaluation assesses how well the model can maintain its segmentation accuracy when the input image undergoes translation transformations, which helps ensure the model's reliability in real-world scenarios: the spine may not be in the center of the image. We limit the translations up to 40 pixels because a greater shift would cause objects to cross or go out of the image frame boundary.

Each model was trained on a Nvidia A6000 GPU with 48GB VRAM. During the training process, a batch size of 6 was used for most of the models, except for training MedT, where a batch size of 2 was utilized due to its large model size. The Adam optimizer with an initial learning rate of 0.0001 was employed for model optimization. A low learning rate is generally preferred to ensure stable convergence during training. Although a low learning rate might slow down the convergence process, it helps avoid convergence failures. In our experiments, each model was trained for approximately 1000 epochs. This choice was made because after 1000 epochs, no further improvement was observed in the performance of the models. Additionally, gradient clipping is applied during training to prevent potentially large gradients from causing instability in the learning process. We performed model selection based on the performance on the validation set.

*4.3. Results of Experiment-A with the original training set*

In experiment-A, the 16 models were trained on the original training set with elastic deformation and random-shift, and then the models were evaluated on both the original test set and the augmented test set to measure instance segmentation accuracy and translation robustness. Figure 6 displays the performance of the top 5 models, and our SymTC performs the best.

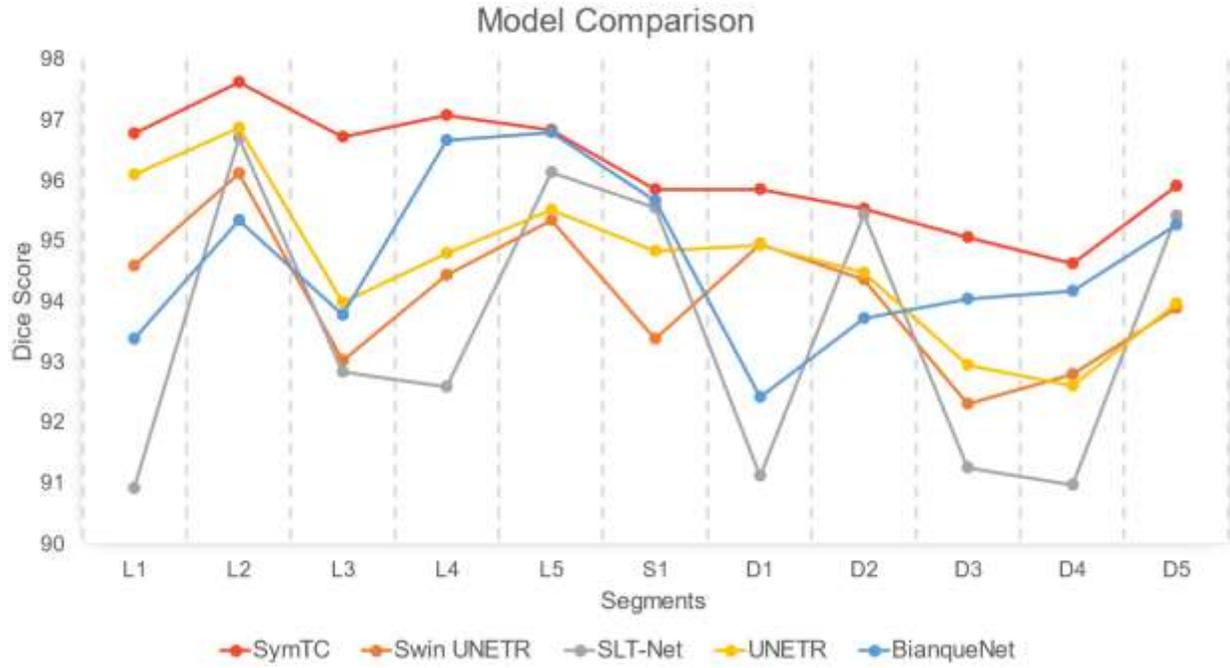

**Fig. 6.** Top 5 Model Comparison Results (DSC) on the augmented test set.

*4.3.1. Instance Segmentation Evaluation*

Table 2 summarizes the instance segmentation results for vertebrae bodies (VB) and intervertebral discs (IVD) in terms of the DSC on the original test set consisting of 20 samples. For better clarity and ease of understanding, we have converted DSC into percentage ratios between 0 and 100%. The results show that our proposed network SymTC outperforms the other 15 models on average. SymTC stands out by achieving the highest DSC for vertebrae bodies L1, L3, L4 and intervertebral discs D1, D2, D3. SymTC along with other Transformer-based models surpasses the CNN-only models.

**Table 2.** DSC (higher is better) of each model on the original test set.

|  | L1 | L2 | L3 | L4 | L5 | S1 | D1 | D2 | D3 | D4 | D5 | Average |
|---|---|---|---|---|---|---|---|---|---|---|---|---|
| SymTC | **95.598** ±1.791 | 97.010 ±1.655 | **95.506** ±6.004 | **96.531** ±0.959 | 95.319 ±4.030 | 93.706 ±4.263 | **94.851** ±2.459 | **94.730** ±4.371 | **94.568** ±3.167 | 93.301 ±3.240 | 92.776 ±9.310 | **94.900** ±2.095 |
| Swin UNETR | 93.126 ±7.356 | 96.813 ±1.641 | 93.874 ±11.389 | 95.805 ±3.189 | 94.380 ±4.816 | 92.194 ±6.449 | 92.757 ±4.568 | 94.299 ±3.519 | 92.821 ±6.563 | 89.777 ±7.645 | 91.077 ±10.179 | 93.357 ±4.098 |
| SLT-Net | 91.967 ±6.991 | 97.052 ±0.902 | 94.328 ±10.261 | 93.619 ±13.504 | 95.763 ±2.553 | 93.805 ±1.815 | 92.990 ±4.893 | 92.938 ±6.914 | 90.282 ±16.896 | 91.481 ±5.639 | 92.589 ±6.048 | 93.347 ±5.886 |
| UNETR | 94.355 ±3.131 | 96.795 ±1.709 | 91.900 ±18.125 | 95.546 ±3.196 | 94.099 ±3.814 | 92.988 ±4.539 | 93.261 ±3.878 | 93.554 ±4.254 | 90.931 ±9.003 | 90.435 ±5.678 | 91.446 ±4.198 | 93.210 ±4.333 |

| | L1 | L2 | L3 | L4 | L5 | S1 | D1 | D2 | D3 | D4 | D5 | Average |
|---|---|---|---|---|---|---|---|---|---|---|---|---|
| BianqueNet | 89.330 ±21.689 | 93.332 ±16.860 | 91.819 ±18.667 | 95.245 ±6.709 | **96.011** ±2.155 | **94.193** ±5.037 | 90.069 ±21.017 | 92.566 ±9.930 | 93.635 ±4.852 | **93.539** ±3.373 | **94.326** ±4.325 | 93.097 ±2.840 |
| Inception-SwinUnet | 91.397 ±19.415 | **97.335** ±0.812 | 91.926 ±20.366 | 93.837 ±12.637 | 91.156 ±18.678 | 91.592 ±8.410 | 93.533 ±4.614 | 93.281 ±7.055 | 89.739 ±20.753 | 87.727 ±14.249 | 91.233 ±11.216 | 92.069 ±9.514 |
| HSNet | 94.240 ±12.384 | 94.116 ±10.544 | 90.123 ±22.355 | 91.649 ±21.243 | 93.456 ±11.688 | 92.662 ±9.901 | 92.948 ±7.903 | 88.941 ±21.43 | 89.008 ±21.266 | 89.158 ±17.099 | 91.986 ±12.085 | 91.662 ±13.085 |
| Swin-Unet | 90.839 ±20.912 | 93.514 ±16.261 | 91.781 ±21.085 | 91.781 ±21.08 | 90.076 ±21.037 | 92.197 ±6.612 | 91.638 ±11.014 | 90.000 ±20.632 | 89.755 ±20.720 | 87.864 ±20.629 | 88.554 ±16.934 | 90.727 ±16.984 |
| UNeXt | 92.354 ±15.512 | 95.835 ±4.808 | 91.220 ±21.036 | 89.282 ±21.93 | 88.818 ±18.828 | 87.325 ±14.428 | 92.082 ±9.072 | 91.271 ±16.774 | 87.280 ±21.834 | 85.790 ±19.574 | 89.065 ±17.579 | 90.029 ±12.849 |
| TransUNet | 92.245 ±9.199 | 87.526 ±25.59 | 88.178 ±24.371 | 91.605 ±21.122 | 91.219 ±21.025 | 89.705 ±20.496 | 81.444 ±32.928 | 86.309 ±24.176 | 88.812 ±21.076 | 88.060 ±20.596 | 90.294 ±20.782 | 88.672 ±20.293 |
| MedT | 88.707 ±20.984 | 94.461 ±8.57 | 91.011 ±20.989 | 90.099 ±20.981 | 87.121 ±25.111 | 87.183 ±15.317 | 87.639 ±20.944 | 89.689 ±19.384 | 87.318 ±20.279 | 82.663 ±23.855 | 85.180 ±22.131 | 88.361 ±18.641 |
| UTNet | 81.322 ±33.18 | 82.886 ±31.442 | 85.159 ±29.06 | 88.802 ±24.539 | 90.776 ±20.988 | 87.080 ±14.823 | 77.511 ±32.691 | 78.711 ±32.650 | 83.856 ±28.478 | 85.608 ±21.674 | 89.312 ±18.307 | 84.639 ±23.716 |
| UCTransNet | 75.346 ±31.945 | 78.997 ±28.28 | 81.862 ±26.04 | 80.907 ±29.371 | 86.107 ±22.629 | 86.968 ±20.202 | 80.058 ±25.635 | 80.407 ±22.833 | 76.787 ±30.266 | 81.144 ±25.322 | 88.063 ±19.596 | 81.513 ±20.303 |
| Attention U-Net | 74.333 ±35.642 | 74.288 ±33.164 | 68.732 ±36.905 | 76.499 ±29.188 | 86.964 ±22.583 | 85.474 ±25.111 | 73.700 ±37.113 | 68.943 ±37.476 | 68.092 ±35.719 | 80.147 ±25.685 | 87.694 ±21.912 | 76.806 ±25.335 |
| UNet++ | 68.734 ±40.68 | 66.998 ±35.694 | 74.033 ±27.911 | 71.468 ±40.229 | 81.297 ±31.709 | 83.980 ±27.527 | 65.450 ±39.937 | 71.473 ±28.932 | 67.208 ±40.117 | 79.834 ±28.708 | 84.529 ±26.387 | 74.091 ±26.461 |
| MultiResUNet | 77.421 ±31.596 | 74.309 ±29.915 | 66.317 ±34.739 | 66.828 ±35.945 | 76.870 ±35.698 | 84.644 ±22.624 | 75.710 ±33.508 | 70.025 ±27.873 | 58.563 ±39.241 | 72.553 ±33.342 | 75.217 ±31.98 | 72.587 ±26.547 |

Table 3 shows the instance segmentation results measured by 95% Hausdorff distance (HD95) on the original test set. The results show that our proposed network SymTC outperforms the other 15 models on average, and it achieves the lowest HD95 for L1, L3, L4, D1, D2, D3, and D5.

**Table 3.** HD95 (lower is better) of each model on the original test set.

| | L1 | L2 | L3 | L4 | L5 | S1 | D1 | D2 | D3 | D4 | D5 | Average |
|---|---|---|---|---|---|---|---|---|---|---|---|---|
| SymTC | **1.93** ±1.829 | 2.006 ±3.88 | 1.949 ±2.183 | **1.64** ±0.591 | 2.553 ±2.47 | 5.1 ±6.993 | **1.134** ±0.449 | **1.626** ±2.227 | **1.491** ±0.711 | 2.118 ±1.234 | **2.778** ±3.973 | **2.211** ±1.187 |
| Swin UNETR | 8.594 ±12.047 | 11.024 ±33.718 | 2.992 ±3.483 | 3.147 ±6.521 | 3.759 ±5.905 | 9.454 ±13.059 | 19.057 ±41.678 | 1.894 ±2.472 | 3.466 ±5.801 | 3.814 ±3.828 | 8.128 ±12.917 | 6.848 ±10.137 |
| SLT-Net | 13.716 ±16.122 | 1.332 ±0.64 | 2.371 ±3.562 | 2.778 ±5.493 | 2.363 ±2.252 | 8.11 ±17.93 | 2.348 ±3.912 | 2.004 ±2.338 | 3.118 ±5.032 | 2.865 ±2.174 | 4.762 ±12.584 | 4.161 ±4.089 |
| UNETR | 3.744 ±6.201 | 1.512 ±1.261 | 3.679 ±7.65 | 2.237 ±1.722 | 5.927 ±6.302 | **4.724** ±5.909 | 2.442 ±3.25 | 1.704 ±1.3 | 3.589 ±4.982 | 3.277 ±2.968 | 7.517 ±8.093 | 3.668 ±2.34 |
| BianqueNet | 5.056 ±10.105 | 7.158 ±15.206 | 3.624 ±7.993 | 3.375 ±7.988 | **2.290** ±2.285 | 5.308 ±9.319 | 4.278 ±9.886 | 5.106 ±10.580 | 3.138 ±7.382 | 5.857 ±12.764 | 3.713 ±6.602 | 4.446 ±9.739 |
| Inception-SwinUnet | 3.378 ±7.275 | **1.197** ±0.555 | 3.5 ±8.067 | 3.033 ±6.334 | 3.362 ±3.809 | 6.39 ±10.21 | 7.53 ±25.967 | 2.051 ±3.129 | 1.59 ±0.876 | 3.957 ±4.714 | 5.715 ±9.957 | 3.791 ±4.134 |
| HSNet | 3.006 ±6.714 | 3.634 ±8.238 | 4.121 ±8.874 | 3.94 ±9.34 | 5.547 ±11.36 | 5.196 ±10.482 | 2.869 ±7.442 | 5.378 ±11.527 | 4.145 ±9.551 | 5.966 ±11.862 | 5.045 ±11.035 | 4.441 ±7.376 |
| Swin-Unet | 2.131 ±3.459 | 2.226 ±4.686 | 1.753 ±1.591 | 1.971 ±2.484 | 2.767 ±2.596 | 4.732 ±7.47 | 2.221 ±3.907 | 2.937 ±6.313 | 1.679 ±1.159 | **2.08** ±1.308 | 6.339 ±10.19 | 2.803 ±2.24 |
| UNeXt | 2.805 ±4.959 | 2.016 ±2.768 | **1.933** ±1.542 | 7.259 ±19.018 | 13.485 ±17.479 | 10.8 ±15.286 | 5.98 ±14.131 | 2.536 ±4.028 | 5.69 ±15.364 | 7.292 ±13.081 | 11.01 ±15.926 | 6.437 ±8.162 |
| TransUNet | 7.727 ±15.874 | 8.556 ±20.98 | 9.261 ±19.824 | 3.836 ±9.163 | 4.372 ±9.063 | 7.595 ±12.805 | 8.968 ±22.198 | 8.662 ±19.145 | 3.876 ±9.056 | 5.176 ±10.127 | 4.15 ±9.494 | 6.562 ±11.646 |
| MedT | 8.665 ±14.016 | 2.557 ±2.564 | 6.168 ±11.39 | 2.386 ±2.751 | 7.056 ±10.149 | 10.406 ±10.254 | 5.684 ±10.784 | 7.024 ±11.272 | 63.762 ±83.727 | 5.359 ±8.133 | 18.524 ±33.337 | 12.508 ±7.994 |
| UTNet | 6.897 ±12.365 | 7.209 ±12.726 | 10.986 ±16.485 | 5.092 ±10.427 | 8.523 ±13.468 | 20.203 ±35.303 | 7.421 ±13.484 | 13.133 ±18.251 | 11.486 ±17.583 | 10.463 ±17.173 | 8.903 ±13.608 | 10.029 ±12.359 |
| UCTransNet | 11.554 ±13.053 | 21.176 ±20.644 | 30.833 ±45.698 | 12.752 ±15.773 | 11.034 ±15.015 | 9.536 ±15.319 | 20.424 ±32.647 | 23.576 ±27.95 | 20.578 ±22.595 | 14.814 ±16.837 | 7.444 ±13.146 | 16.702 ±12.286 |
| Attention U-Net | 9.394 ±12.866 | 13.821 ±17.43 | 16.946 ±18.222 | 21.225 ±32.217 | 10.944 ±15.876 | 7.474 ±13.279 | 13.878 ±18.345 | 15.805 ±19.732 | 28.632 ±35.708 | 10.027 ±15.161 | 7.124 ±13.263 | 14.116 ±13.022 |
| UNet++ | 6.733 ±11.95 | 16.872 ±17.754 | 22.11 ±20.232 | 15.901 ±19.52 | 12.515 ±16.298 | 9.518 ±14.037 | 9.862 ±14.455 | 21.599 ±20.421 | 19.863 ±26.463 | 10.369 ±16.036 | 9.929 ±15.288 | 14.116 ±11.826 |
| MultiResUNet | 8.841 ±11.614 | 17.292 ±16.037 | 19.035 ±17.013 | 15.814 ±15.802 | 9.93 ±13.24 | 9.582 ±13.03 | 9.63 ±14.817 | 18.331 ±18.619 | 22.223 ±19.32 | 10.918 ±15.737 | 10.632 ±14.034 | 13.839 ±11.524 |

We also assessed the segmentation performance of all models on the augmented test set consisting of 2500 samples. Table 4 (DSC) and Table 5 (HD95) summarize the instance segmentation performance of each model evaluated on the augmented test set. Our proposed network SymTC outperforms the other 15 models on average. SymTC has the highest average DSC for L1, L3, L4, L5, D1, D2, and D3, and attains the lowest HD95 for L1, L3, L4, D1, D2, D3, and D4.

**Table 4.** DSC (higher is better) of each model on the augmented test set.

|  | L1 | L2 | L3 | L4 | L5 | S1 | D1 | D2 | D3 | D4 | D5 | Average |
|---|---|---|---|---|---|---|---|---|---|---|---|---|
| SymTC | **95.718** ±2.951 | 96.956 ±2.787 | **95.265** ±7.417 | **96.606** ±1.418 | 96.153 ±1.89 | 94.448 ±3.89 | **94.913** ±2.55 | 94.827 ±4.502 | **94.427** ±3.683 | 93.334 ±4.155 | 94.231 ±3.729 | **95.171** ±2.144 |
| Swin UNETR | 93.108 ±8.399 | 96.337 ±3.289 | 93.286 ±14.544 | 95.270 ±5.054 | 94.695 ±4.905 | 93.072 ±5.52 | 93.127 ±5.447 | 94.191 ±4.227 | 91.970 ±9.192 | 90.438 ±6.799 | 91.308 ±6.215 | 93.346 ±4.983 |
| SLT-Net | 91.925 ±11.161 | 96.736 ±2.401 | 93.889 ±12.742 | 92.538 ±18.37 | 94.557 ±8.863 | 92.437 ±8.045 | 92.712 ±7.199 | 93.833 ±4.888 | 89.858 ±18.676 | 90.419 ±11.831 | 91.149 ±11.399 | 92.732 ±9.07 |
| UNETR | 93.839 ±6.152 | 96.260 ±3.454 | 91.773 ±18.475 | 94.984 ±5.13 | 94.052 ±5.003 | 89.817 ±10.051 | 93.039 ±3.998 | 93.865 ±4.332 | 90.486 ±9.81 | 90.317 ±5.791 | 89.976 ±6.801 | 92.583 ±5.384 |
| BianqueNet | 92.103 ±16.133 | 95.329 ±10.126 | 94.670 ±10.097 | 95.486 ±7.397 | 96.003 ±2.653 | 95.191 ±2.992 | 92.275 ±13.814 | 93.724 ±7.138 | 93.752 ±6.614 | **93.384** ±5.339 | 94.187 ±4.917 | 94.191 ±8.996 |
| Inception-SwinUnet | 91.988 ±18.095 | **96.977** ±3.847 | 91.923 ±20.632 | 92.382 ±18.076 | 94.237 ±8.303 | 92.904 ±9.058 | 93.660 ±6.763 | 93.689 ±6.904 | 89.551 ±20.673 | 88.781 ±14.933 | 91.991 ±7.246 | 92.553 ±10.665 |
| HSNet | 94.628 ±14.373 | 95.868 ±10.502 | 92.061 ±18.001 | 91.512 ±21.12 | 95.849 ±8.781 | **95.682** ±3.865 | 93.117 ±13.282 | 93.874 ±10.222 | 89.820 ±19.811 | 89.645 ±17.916 | **94.570** ±4.643 | 93.196 ±9.196 |
| Swin-Unet | 91.328 ±18.901 | 93.299 ±17.946 | 91.703 ±21.094 | 91.669 ±21.131 | 91.198 ±18.97 | 91.366 ±12.268 | 91.283 ±14.972 | 90.320 ±20.105 | 89.189 ±20.739 | 87.815 ±20.844 | 90.297 ±13.743 | 90.861 ±17.347 |
| UNeXt | 91.280 ±16.796 | 95.036 ±8.14 | 91.013 ±21.094 | 88.953 ±23.119 | 91.170 ±13.263 | 89.812 ±12.544 | 91.255 ±11.141 | 92.186 ±12.876 | 87.131 ±22.224 | 87.645 ±15.647 | 90.288 ±10.6 | 90.525 ±13.111 |
| TransUNet | 92.319 ±16.2 | 93.879 ±16.636 | 92.586 ±17.395 | 94.676 ±10.604 | 94.741 ±7.245 | 94.003 ±5.979 | 90.806 ±18.041 | 91.343 ±16.735 | 90.749 ±16.481 | 91.550 ±10.169 | 92.894 ±6.709 | 92.686 ±10.357 |
| MedT | 86.861 ±23.723 | 92.629 ±15.296 | 91.421 ±19.013 | 89.039 ±21.809 | 88.147 ±20.587 | 84.398 ±18.84 | 86.124 ±22.55 | 89.722 ±19.247 | 86.538 ±20.526 | 83.335 ±23.549 | 83.075 ±21.478 | 87.390 ±18.852 |
| UTNet | 82.709 ±28.31 | 80.782 ±30.95 | 84.073 ±26.989 | 85.725 ±27.893 | 91.096 ±16.215 | 91.306 ±9.014 | 76.822 ±30.988 | 80.621 ±28.224 | 83.916 ±27.028 | 85.401 ±20.584 | 90.734 ±11.839 | 84.835 ±19.788 |
| UCTransNet | 72.343 ±34.307 | 76.324 ±29.91 | 81.393 ±24.405 | 79.606 ±28.661 | 87.287 ±18.466 | 92.642 ±8.999 | 73.083 ±32.634 | 81.171 ±23.118 | 74.872 ±29.442 | 79.360 ±23.231 | 89.673 ±13.893 | 80.725 ±18.567 |
| Attention U-Net | 68.915 ±36.382 | 69.948 ±35.755 | 65.592 ±34.768 | 75.681 ±28.502 | 86.721 ±22.809 | 93.227 ±8.204 | 66.819 ±38.116 | 63.876 ±37.865 | 66.104 ±32.424 | 75.905 ±28.8 | 91.537 ±11.763 | 74.939 ±20.816 |
| UNet++ | 65.096 ±40.792 | 69.498 ±36.367 | 69.993 ±31.743 | 66.925 ±37.244 | 81.392 ±28.029 | 90.460 ±13.191 | 64.565 ±40.33 | 72.047 ±31.75 | 63.061 ±38.234 | 72.931 ±31.167 | 87.159 ±20.266 | 73.012 ±23.317 |
| MultiResUNet | 73.766 ±33.559 | 72.475 ±32.877 | 70.752 ±32.574 | 70.171 ±35.186 | 78.962 ±32.165 | 87.809 ±18.254 | 68.730 ±36.759 | 69.854 ±34.213 | 68.738 ±34.452 | 72.479 ±33.345 | 79.657 ±28.326 | 73.945 ±25.091 |

**Table 5.** HD95 value (lower is better) of each model on the augmented test set.

|  | L1 | L2 | L3 | L4 | L5 | S1 | D1 | D2 | D3 | D4 | D5 | Average |
|---|---|---|---|---|---|---|---|---|---|---|---|---|
| SymTC | **2.188** ±3.968 | 1.984 ±3.787 | **2.047** ±2.456 | **1.741** ±1.435 | 2.234 ±2.999 | 4.668 ±10.907 | **1.566** ±3.679 | **1.79** ±3.219 | **1.739** ±2.695 | **2.204** ±1.461 | 4.064 ±9.357 | **2.384** ±2.108 |
| Swin UNETR | 11.353 ±33.189 | 9.468 ±31.478 | 7.005 ±23.16 | 5.783 ±17.828 | 5.375 ±11.093 | 16.07 ±35.564 | 13.829 ±40.509 | 3.708 ±15.61 | 6.191 ±18.392 | 4.775 ±11.302 | 11.15 ±17.265 | 8.609 ±15.102 |
| SLT-Net | 8.464 ±17.817 | 2.537 ±9.812 | 3.463 ±9.614 | 3.078 ±8.708 | 3.766 ±9.767 | 5.409 ±15.909 | 3.96 ±15.752 | 2.376 ±5.825 | 3.835 ±13.919 | 4.023 ±10.257 | 5.751 ±11.967 | 4.242 ±7.772 |
| UNETR | 5.873 ±10.615 | 2.63 ±6.155 | 4.111 ±7.566 | 2.506 ±3.004 | 5.014 ±7.516 | 10.871 ±19.237 | 2.772 ±4.803 | 2.103 ±3.727 | 4.111 ±6.555 | 3.735 ±4.708 | 6.404 ±8.943 | 4.557 ±4.305 |
| BianqueNet | 5.555 ±10.365 | 5.767 ±13.823 | 3.398 ±8.156 | 3.173 ±6.849 | 3.318 ±7.518 | 3.540 ±11.313 | 5.303 ±11.274 | 5.818 ±15.604 | 2.231 ±4.499 | 2.868 ±6.468 | 3.384 ±7.982 | 4.032 ±10.045 |
| Inception-SwinUnet | 3.41 ±10.003 | **1.529** ±5.754 | 3.685 ±10.956 | 3.125 ±7.22 | 4.276 ±10.702 | 5.534 ±13.91 | 1.677 ±3.326 | 2.424 ±7.185 | 3.671 ±9.82 | 3.923 ±6.581 | 6.093 ±11.203 | 3.577 ±5.158 |
| HSNet | 2.575 ±6.116 | 2.72 ±9.091 | 3.826 ±9.386 | 2.955 ±7.523 | **1.958** ±5.101 | **2.37** ±4.584 | 2.165 ±6.245 | 2.778 ±7.9 | 3.13 ±8.164 | 2.894 ±6.34 | **2.258** ±6.682 | 2.693 ±4.533 |
| Swin-Unet | 3.718 ±12.477 | 2.281 ±9.182 | 2.092 ±5.183 | 2.044 ±5.193 | 4.841 ±11.639 | 5.539 ±9.335 | 2.601 ±10.854 | 2.392 ±9.554 | 2.103 ±4.736 | 2.735 ±5.601 | 6.01 ±10.775 | 3.305 ±4.993 |
| UNeXt | 7.212 ±18.727 | 4.233 ±12.846 | 4.913 ±16.54 | 7.77 ±17.801 | 11.646 ±17.788 | 9.462 ±21.051 | 8.457 ±20.29 | 3.848 ±10.381 | 7.327 ±19.721 | 7.806 ±12.668 | 11.651 ±16.133 | 7.666 ±9.932 |
| TransUNet | 6.245 ±18.622 | 6.008 ±18.599 | 6.191 ±17.579 | 5.246 ±18.663 | 6.385 ±20.503 | 7.016 ±19.462 | 5.417 ±18.977 | 8.688 ±23.985 | 7.87 ±21.105 | 5.241 ±14.506 | 9.589 ±24.032 | 6.718 ±13.318 |
| MedT | 8.115 ±12.876 | 3.878 ±6.953 | 4.385 ±7.671 | 5.283 ±9.288 | 10.009 ±13.817 | 13.935 ±16.408 | 6.139 ±11.146 | 4.428 ±9.506 | 72.075 ±82.33 | 8.635 ±16.218 | 12.997 ±17.864 | 13.625 ±9.645 |
| UTNet | 8.667 ±17.117 | 10.592 ±20.449 | 15.484 ±20.453 | 9.202 ±17.838 | 14.237 ±27.825 | 14.724 ±22.491 | 10.674 ±16.985 | 15.102 ±21.452 | 11.572 ±19.287 | 10.649 ±18.77 | 8.581 ±16.592 | 11.771± 11.935 |
| UCTransNet | 14.267 ±20.828 | 21.247 ±25.68 | 23.542 ±31.268 | 20.769 ±31.321 | 12.504 ±16.412 | 9.226 ±22.252 | 21.576 ±32.676 | 27.888 ±31.415 | 28.971 ±34.852 | 17.921 ±25.073 | 10.608 ±19.215 | 18.956 ±13.438 |
| Attention U-Net | 21.033 ±29.85 | 18.984 ±20.704 | 22.082 ±22.065 | 31.271 ±36.969 | 11.244 ±20.511 | 6.783 ±16.671 | 18.986 ±25.984 | 19 ±21.259 | 32.27 ±31.194 | 19.275 ±26.051 | 11.469 ±28.816 | 19.309 ±14.67 |
| UNet++ | 18.011 ±27.449 | 16.192 ±19.325 | 23.953 ±20.546 | 18.749 ±21.889 | 12.423 ±16.826 | 6.939 ±12.271 | 13.378 ±20.277 | 20.232 ±21.096 | 25.36 ±30.082 | 17.679 ±20.881 | 8.728 ±15.29 | 16.513 ±12.019 |
| MultiResUNet | 13.291 ±19.988 | 16.623 ±17.923 | 20.534 ±20.18 | 16.424 ±18.717 | 8.357 ±12.915 | 7.503 ±11.828 | 13.491 ±20.581 | 20.504 ±23.767 | 19.608 ±22.555 | 11.27 ±16.494 | 8.239 ±13.843 | 14.168 ±11.371 |

To further compare the performance, we conducted a Paired t-Test (parametric) and a Kolmogorov-Smirnov test (non-parametric) using the results on the augmented test set with 2500 samples. The null hypothesis is that the paired segmentation performance between SymTC and another model is the same. Given that the p-values of both the parametric and non-parametric statistical tests are less than 0.05, we reject the null hypothesis and conclude that the paired instance segmentation performance between SymTC and each of the other 15 methods is significantly different. We conducted the statistical tests on the augmented test set because conventional inferential statistics may lack validity for small sample sizes ($n < 30$), and larger sample sizes may be necessary in heavy-tailed distributions [84–86].

The results in Tables 2-5 highlight the superior performance and effectiveness of SymTC in consistently achieving top instance segmentation results across both DSC and HD95 evaluation criteria on both the original and augmented test sets. It is also evident that Transformer-based models surpass traditional CNN-only models in instance segmentation performance.

Figure 7 shows segmentation examples of the 16 models. It is evident that SymTC accurately identifies all vertebrae and lumbar discs in the lumbar MR image, whereas other models produce misclassifications or fragmentation errors. For example, BianqueNet exhibits hollow holes in both D3 and D4. TransUnet incorrectly identifies D3 as D2 and exhibits a significant segmentation error in the lower left corner in the MR scan. Because DSC and HD95 capture distinct aspects of segmentation performance, it is important to take both into account. Even though SymTC may have a slightly lower DSC compared to BianqueNet in Figure 7, SymTC achieves a significantly lower HD95 than BianqueNet. As a result, the segmentation results depicted in Figure 7 appear to be more realistic and accurate when using SymTC, highlighting the importance of considering both DSC and HD95 for a comprehensive evaluation. Also seen in Figure 7, many other models produce incorrect predictions around pixels in close proximity to the boundary of two adjacent objects, and SymTC demonstrates superior boundary delineation. Due to our method's consideration of pixel-wise dependencies encoded with the proposed relative position embedding, it demonstrates a more effective learning of these dependencies compared to other methods. This results in our predictions being more accurate, as there are less misclassified pixels near object boundaries.

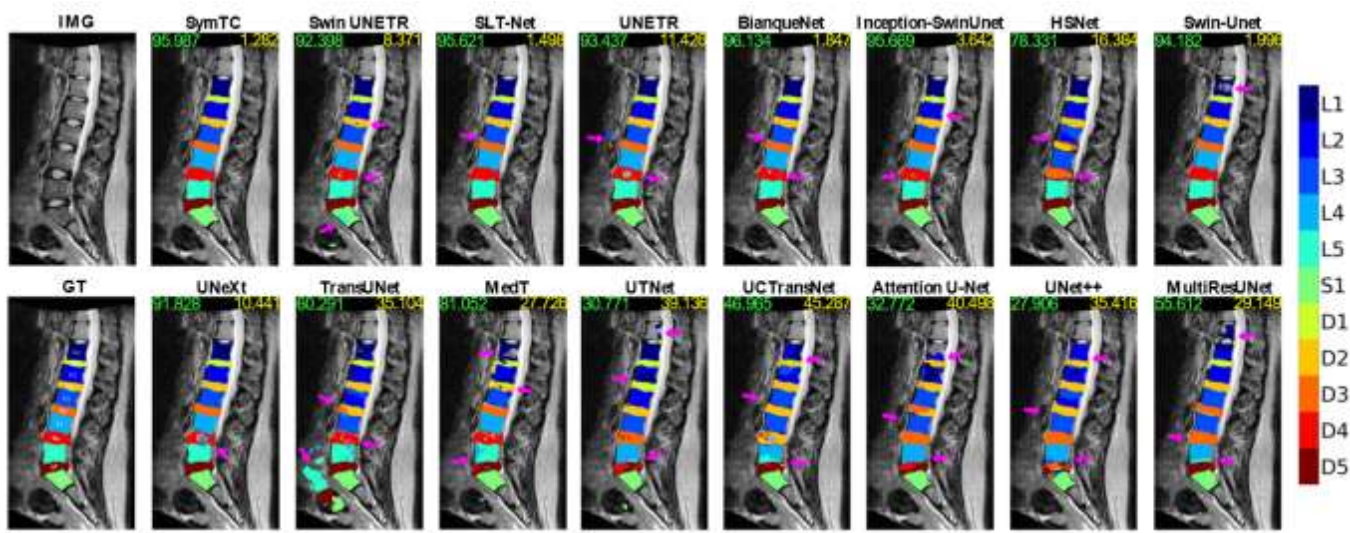

**Fig. 7.** Segmentation examples of the 16 models. IMG is the input image. GT indicates the Ground-Truth annotation. DSC (green color) and HD95 (yellow color) are shown on top corners of each image. The 11 lumbar objects are shown in different colors. Large segmenttion errors are indicated by pink arrows.

*4.3.2. Translation Robustness Evaluation*

We also assess the translation robustness of our SymTC model in comparison to other models. Our objective is to assess the performance of an instance segmentation model when it encounters shifted MR images. This assessment provides insights into a model's ability to generalize and maintain reliability in the real scenario where the overall position of the spine on an image may vary.

Table 6 and Table 7 exhibit each model's resilience in response to image shifts in the horizontal direction and the vertical direction, respectively. SymTC has the highest robustness performance in the horizontal direction, and it also has the highest robustness performance when the vertical translation is constrained within 20 pixels. BianqueNet excels in the case of vertical translation ranging from 20 to 40 pixels. We note the robustness of SymTC will increase when training on the large augmented training set in experiment-B (see Section 4.4.2).

**Table 6.** DSC results (higher is better) of horizontal-shit robustness experiment on the augmented test set. The displacement values are 10, 20, 30, 40 (pixels). An image could be shifted to the left or to the right. The results of left-shift and right-shift with the same displacement are averaged.

|  | 0 | 10 | 20 | 30 | 40 |
|---|---|---|---|---|---|
| SymTC | **95.171** ±2.144 | **95.159** ±2.181 | **95.110** ±2.293 | **95.010** ±2.472 | **94.789** ±2.848 |
| Swin UNETR | 93.346 ±4.983 | 93.297 ±5.113 | 93.123 ±5.545 | 92.745 ±6.252 | 92.098 ±7.266 |
| SLT-Net | 92.732 ±9.070 | 92.752 ±8.977 | 92.626 ±9.404 | 92.346 ±10.338 | 92.067 ±11.260 |
| UNETR | 92.583 ±5.384 | 92.390 ±5.708 | 91.764 ±6.634 | 89.973 ±8.216 | 85.782 ±10.576 |
| BianqueNet | 94.191 ±5.633 | 94.181 ±5.660 | 94.158 ±5.706 | 94.158 ±5.684 | 94.137 ±5.669 |
| Inception-SwinUnet | 92.553 ±10.665 | 92.447 ±11.029 | 92.176 ±11.950 | 91.827 ±12.897 | 91.389 ±13.927 |
| HSNet | 93.330 ±9.196 | 93.366 ±8.990 | 93.407 ±8.908 | 93.347 ±9.076 | 93.333 ±8.993 |
| Swin-Unet | 90.861 ±17.347 | 90.717 ±17.550 | 90.353 ±17.975 | 89.849 ±18.228 | 88.785 ±18.755 |
| UneXt | 90.525 ±13.111 | 90.466 ±13.164 | 90.450 ±13.175 | 90.484 ±13.166 | 90.415 ±13.239 |
| TransUNet | 92.686 ±10.357 | 92.679 ±10.362 | 92.700 ±10.191 | 92.721 ±10.089 | 92.782 ±9.797 |
| MedT | 87.390 ±18.852 | 87.295 ±18.962 | 87.011 ±19.186 | 86.485 ±19.497 | 85.693 ±19.878 |
| UTNet | 84.835 ±19.788 | 84.786 ±19.748 | 84.292 ±19.990 | 83.662 ±20.260 | 83.643 ±20.269 |
| UCTransNet | 80.725 ±18.567 | 80.648 ±18.658 | 80.381 ±18.740 | 79.983 ±18.880 | 79.339 ±19.091 |
| Attention U-Net | 74.939 ±20.816 | 74.857 ±20.886 | 74.892 ±20.875 | 74.893 ±20.883 | 74.880 ±20.850 |
| Unet++ | 73.012 ±23.317 | 72.886 ±23.393 | 72.934 ±23.352 | 72.972 ±23.349 | 72.892 ±23.347 |
| MultiResUNet | 73.945 ±25.091 | 73.935 ±25.078 | 73.892 ±25.150 | 73.958 ±25.086 | 73.962 ±25.086 |

**Table 7.** DSC results (higher is better) of vertical-shit robustness experiment on the augmented test set. The displacement values are 10, 20, 30, 40 (pixels). An image could be shifted up or down. The results of up-shift and down-shift with the same displacement are averaged.

| | 0 | 10 | 20 | 30 | 40 |
|---|---|---|---|---|---|
| SymTC | **95.171** ±2.144 | **95.075** ±2.918 | **94.299** ±6.097 | 88.772 ±15.911 | 65.746 ±28.109 |
| Swin UNETR | 93.346 ±4.983 | 93.349 ±5.195 | 91.459 ±9.470 | 74.140 ±23.991 | 40.481 ±21.560 |
| SLT-Net | 92.732 ±9.070 | 92.545 ±9.744 | 91.022 ±11.802 | 70.455 ±25.555 | 38.540 ±17.218 |
| UNETR | 92.583 ±5.384 | 92.349 ±5.903 | 90.150 ±10.968 | 73.600 ±26.210 | 29.801 ±23.044 |
| BianqueNet | 94.191 ±5.633 | 94.557 ±4.733 | 94.297 ±5.533 | **93.050** ±8.311 | **89.889** ±12.777 |
| Inception-SwinUnet | 92.553 ±10.665 | 92.535 ±10.533 | 91.177 ±12.514 | 80.521 ±21.338 | 54.437 ±24.638 |
| HSNet | 93.330 ±9.196 | 93.741 ±8.389 | 93.538 ±9.012 | 92.465 ±10.983 | 89.213 ±15.011 |
| Swin-Unet | 90.861 ±17.347 | 90.636 ±17.501 | 89.176 ±18.612 | 73.142 ±26.973 | 38.689 ±21.415 |
| UneXt | 90.525 ±13.111 | 90.632 ±13.205 | 89.622 ±14.364 | 85.245 ±17.722 | 75.723 ±21.031 |
| TransUNet | 92.686 ±10.357 | 92.802 ±10.377 | 92.416 ±10.624 | 90.996 ±11.931 | 85.995 ±15.036 |
| MedT | 87.390 ±18.852 | 87.375 ±18.907 | 84.770 ±19.918 | 76.460 ±24.128 | 59.687 ±27.708 |
| UTNet | 84.835 ±19.788 | 84.781 ±19.880 | 83.949 ±20.203 | 82.269 ±20.667 | 79.665 ±21.264 |
| UCTransNet | 80.725 ±18.567 | 80.316 ±18.875 | 78.338 ±19.653 | 75.076 ±20.717 | 71.053 ±21.402 |
| Attention U-Net | 74.939 ±20.816 | 74.997 ±20.791 | 75.251 ±20.859 | 75.328 ±20.816 | 75.199 ±20.960 |
| Unet++ | 73.012 ±23.317 | 73.030 ±23.300 | 73.091 ±23.300 | 73.190 ±23.302 | 72.890 ±23.295 |
| MultiResUNet | 73.945 ±25.091 | 74.170 ±25.118 | 74.513 ±25.140 | 74.424 ±25.160 | 74.223 ±25.127 |

While the CNN-only models do not attain high DSC for instance segmentation, these models exhibit a notable degree of stability and consistency in the translation robustness experiments. This is attributed to the translation invariance property inherent to CNNs, which makes these models less susceptible to translations in the images. Conversely, Transformer-based models necessitate processing global information, which can pose challenges in dealing with positional shifts. Unlike CNN-only models, which benefit from their translation invariance property, Transformer-based models depend on capturing contextual relationships and dependencies among various elements within the input data. Consequently, the introduction of positional shifts in the input can disrupt the position information in Transformers, leading to potential impacts on the segmentation output. This underscores the limitation of Transformer-based models in the context of instance segmentation tasks and emphasizes the necessity to effectively preserve the relative position information in the input images.

We also observe that, when comparing vertical robustness experiments to horizontal robustness experiments, all models exhibit subpar performance. The models are more sensitive to shifts in the vertical direction than in the horizontal direction. This is due to the inherent property of spine MRI scans, wherein the texture and structure along the vertical direction provide more informative features than those along the horizontal direction.

### 4.4. Results of Experiment-B with the augmented training set

In this section, we show the advantages of our data augmentation method in Section 3.4. The top five models in Table 2 are SymTC, Swin UNETR, SLT-Net, UNETR, and BianqueNet. In the experiment-B, each of the five models was trained from scratch using the augmented training set consisting of 7000 samples. Model evaluations were conducted on both the original test set (see Table 8 and Table 9) and the augmented test set (see Table 10 and Table 11). The results show that the segmentation performances of most models are improved, suggesting that the augmented data can enhance the generalization and overall performance of a model. SymTC exhibits the highest average instance segmentation performance compared to the other four models.

It Is noteworthy that training Transformer-based models with limited data is challenging. Nevertheless, the new data augmentation method effectively tackles the challenge of data scarcity with the support of SSM and biomechanics. Our data augmentation can generate synthetic images that closely resemble real data, effectively enhancing the model training process. This approach assists in augmenting the available data, enabling Transformer models to learn from a more diverse and representative dataset. The augmented/synthesized datasets can be made publicly available without any concerns related to medical data privacy.

By using the instance segmentation evaluation and the additional translation robustness assessment, a comprehensive evaluation of each model's performance is achieved. This ensures that models excel not only instance segmentation ability but also maintain their accuracy under realistic and varied conditions.

#### 4.4.1. Instance Segmentation Evaluation

Table 8 (DSC) and Table 9 (HD95) present the instance segmentation evaluations of the five models on the original test set. It is important to note that the three models have shown improvement in both DSC and HD95. This indicates that our augmented training dataset has indeed enhanced the generalization and performance of the models on previously unseen data. SymTC, in particular, continues to exhibit the best average instance segmentation performance.

**Table 8.** DSC results (higher is better) of each model on the original test set. The "change" is the DSC difference when training on the augmented training set as opposed to training on the original training set.

| Network | L1 | L2 | L3 | L4 | L5 | S1 | D1 | D2 | D3 | D4 | D5 | Average | Change |
|---|---|---|---|---|---|---|---|---|---|---|---|---|---|
| SymTC | **96.816** ±0.751 | **97.418** ±0.793 | **96.350** ±3.726 | **97.025** ±0.911 | **96.728** ±1.917 | 94.089 ±5.211 | **95.449** ±2.457 | 95.004 ±3.991 | **94.747** ±3.819 | 93.640 ±3.778 | **95.398** ±1.742 | **95.697** ±3.253 | **+0.797** |
| Swin UNETR | 95.557 ±2.545 | 97.156 ±1.916 | 94.463 ±10.808 | 95.426 ±6.947 | 94.238 ±8.987 | 91.819 ±8.224 | 94.585 ±3.476 | 93.744 ±5.749 | 93.511 ±7.831 | 89.653 ±13.524 | 91.254 ±10.718 | 93.764 ±8.387 | **+2.687** |
| SLT-Net | 91.590 ±21.053 | 96.189 ±5.735 | 92.721 ±18.353 | 89.381 ±23.918 | 91.403 ±21.091 | 90.406 ±20.637 | 90.319 ±20.655 | **95.059** ±3.031 | 92.061 ±12.894 | 85.353 ±24.918 | 90.444 ±20.790 | 91.357 ±19.053 | **-1.990** |
| UNETR | 95.874 ±2.757 | 96.896 ±1.548 | 93.962 ±11.395 | 95.139 ±7.462 | 95.622 ±2.924 | **94.609** ±2.357 | 94.566 ±2.761 | 94.035 ±4.634 | 93.233 ±7.606 | 91.491 ±6.922 | 93.619 ±4.509 | 94.458 ±5.924 | **+1.248** |
| BianqueNet | 87.489 ±27.764 | 89.546 ±24.220 | 90.307 ±19.149 | 93.448 ±11.174 | 94.369 ±6.939 | 93.031 ±6.960 | 85.923 ±28.719 | 88.646 ±9.390 | 91.167 ±11.079 | 87.369 ±17.608 | 91.668 ±11.856 | 90.269 ±18.575 | **-2.828** |

**Table 9.** HD95 results (lower is better) of each model on the original test set. The "change" is the HD95 difference when training on the augmented training set as opposed to training on the original training set.

| Network | L1 | L2 | L3 | L4 | L5 | S1 | D1 | D2 | D3 | D4 | D5 | Average | Change |
|---|---|---|---|---|---|---|---|---|---|---|---|---|---|
| SymTC | **1.280** ±0.489 | 1.933 ±3.834 | **1.784** ±2.279 | **1.449** ±0.812 | **1.721** ±1.719 | 5.426 ±11.004 | **1.057** ±0.417 | 1.558 ±2.270 | **1.429** ±0.916 | **2.043** ±1.488 | 3.462 ±6.542 | **2.104** ±4.392 | **-0.107** |
| Swin UNETR | 25.542 ±51.669 | 4.810 ±16.491 | 10.771 ±40.426 | 9.623 ±35.968 | 2.767 ±3.436 | 17.774 ±37.188 | 7.049 ±21.414 | 9.973 ±35.002 | 1.481 ±1.111 | 2.975 ±3.206 | 11.992 ±16.526 | 9.524 ±29.767 | **+2.676** |
| SLT-Net | 3.727 ±7.864 | **1.627** ±2.401 | 3.233 ±7.368 | 2.239 ±4.187 | 4.125 ±8.733 | **4.014** ±7.567 | 1.936 ±3.702 | **1.461** ±1.533 | 3.101 ±7.624 | 5.300 ±10.626 | 4.700 ±10.386 | 3.224 ±7.289 | **-0.937** |
| UNETR | 2.333 ±3.715 | 1.715 ±2.224 | 2.753 ±4.274 | 2.525 ±4.489 | 2.623 ±2.789 | 4.075 ±6.766 | 1.380 ±0.580 | 1.922 ±2.882 | 2.279 ±3.703 | 2.551 ±2.389 | **2.256** ±1.511 | 2.401 ±3.641 | **-1.267** |
| BianqueNet | 4.809 ±10.395 | 4.665 ±10.792 | 7.094 ±16.348 | 3.756 ±8.038 | 5.245 ±10.295 | 9.090 ±15.080 | 4.773 ±11.239 | 6.979 ±16.887 | 8.547 ±23.035 | 5.799 ±11.210 | 7.777 ±13.558 | 6.230 ±14.053 | **+1.784** |

Table 10 (DSC) and Table 11 (HD95) present the instance segmentation evaluation results on the augmented test set. It is evident that the segmentation performances of all models have improved by using the augmented training dataset, as indicated by enhancements in both evaluation metrics, except for Swin UNETR on the HD95 metric. The augmented training dataset offers a broader domain, enabling the models to acquire more diverse knowledge and achieve improved segmentation performance. This increased resilience to data variations is a notable benefit of the augmented training dataset. Increasing the complexity of the training dataset helps prevent overfitting and regulating models from memorizing the training. Furthermore, SymTC maintains the best average instance segmentation performance on both evaluation metrics.

**Table 10.** DSC results (higher is better) of each model on the augmented test set. The "change" is the DSC difference when training on the augmented training set as opposed to training on the original training set.

| | L1 | L2 | L3 | L4 | L5 | S1 | D1 | D2 | D3 | D4 | D5 | Average | Change |
|---|---|---|---|---|---|---|---|---|---|---|---|---|---|
| SymTC | **96.767** ±1.020 | **97.618** ±0.766 | **96.723** ±2.621 | **97.081** ±1.120 | **96.825** ±1.389 | **95.855** ±1.912 | **95.862** ±1.873 | **95.541** ±3.147 | **95.054** ±3.446 | **94.624** ±2.479 | **95.912** ±1.390 | **96.169** ±2.277 | **+0.998** |
| Swin UNETR | 94.593 ±8.113 | 96.124 ±7.352 | 93.027 ±17.824 | 94.442 ±12.723 | 95.345 ±6.840 | 93.384 ±6.406 | 94.949 ±3.461 | 94.369 ±7.674 | 92.323 ±13.660 | 92.806 ±5.711 | 93.900 ±4.372 | 94.115 ±9.581 | **+0.769** |
| SLT-Net | 90.928 ±21.398 | 96.708 ±5.097 | 92.836 ±18.462 | 92.604 ±19.937 | 96.143 ±4.279 | 95.545 ±2.669 | 91.135 ±18.062 | 95.435 ±3.448 | 91.254 ±17.130 | 90.982 ±15.676 | 95.427 ±2.972 | 93.545 ±14.102 | **+0.813** |
| UNETR | 96.099 ±2.574 | 96.878 ±2.479 | 93.981 ±11.711 | 94.798 ±8.922 | 95.509 ±3.888 | 94.832 ±2.951 | 94.930 ±2.159 | 94.486 ±5.341 | 92.957 ±8.036 | 92.616 ±4.935 | 93.974 ±3.926 | 94.662 ±6.081 | **+2.059** |
| BianqueNet | 93.381 ±16.166 | 95.339 ±12.916 | 93.786 ±13.760 | 96.670 ±4.308 | 96.789 ±2.243 | 95.666 ±2.302 | 92.427 ±16.211 | 93.723 ±11.633 | 94.040 ±7.633 | 94.184 ±4.318 | 95.276 ±3.268 | 94.662 ±10.231 | **+0.471** |

**Table 11.** HD95 results (lower is better) of each model on the augmented test set. The "change" is the HD95 difference when training on the augmented training set as opposed to training on the original training set.

| | L1 | L2 | L3 | L4 | L5 | S1 | D1 | D2 | D3 | D4 | D5 | Average | Change |
|---|---|---|---|---|---|---|---|---|---|---|---|---|---|
| SymTC | **1.268** ±0.597 | **1.185** ±1.971 | **1.824** ±2.371 | **1.375** ±0.741 | **1.603** ±1.612 | **1.803** ±2.546 | **1.057** ±1.692 | 1.517 ±2.297 | **1.475** ±1.776 | **1.799** ±1.189 | **1.559** ±3.059 | **1.497** ± 1.959 | **-0.887** |
| Swin UNETR | 17.591 ±52.428 | 13.865 ±44.911 | 12.207 ±42.917 | 11.290 ±38.870 | 10.393 ±35.903 | 19.144 ±52.802 | 9.570 ±35.018 | 12.249 ±43.617 | 7.145 ±30.825 | 5.524 ±21.039 | 11.808 ±30.508 | 11.889 ±40.236 | **+3.280** |
| SLT-Net | 3.720 ±8.774 | 1.577 ±3.310 | 3.227 ±7.838 | 1.899 ±5.205 | 2.012 ±2.264 | 1.871 ±2.843 | 2.344 ±6.312 | **1.387** ±2.514 | 3.100 ±8.190 | 2.584 ±4.752 | 1.602 ±2.714 | 2.302 ±5.549 | **-1.940** |
| UNETR | 1.923 ±2.598 | 1.774 ±3.164 | 2.679 ±4.416 | 3.029 ±7.859 | 3.056 ±5.502 | 3.340 ±5.462 | 1.329 ±1.5817 | 1.788 ±2.9369 | 2.348 ±3.860 | 2.640 ±3.108 | 2.455 ±3.079 | 2.396 ±4.341 | **-2.161** |
| BianqueNet | 3.815 ±9.985 | 2.907 ±8.092 | 3.849 ±10.217 | 2.605 ±9.389 | 2.268 ±7.335 | 2.667 ±10.589 | 3.383 ±9.089 | 4.257 ±11.508 | 3.149 ±10.065 | 2.061 ±5.126 | 2.860 ±13.087 | 3.075 ±9.734 | **-0.957** |

We conducted both Paired t-Test (parametric) and Kolmogorov-Smirnov Test (non-parametric) statistical analyses between SymTC and each of the other four methods. The p-values obtained from both statistical tests are less than 0.05, indicating that the average segmentation performance of SymTC is significantly different/better from that of the other models.

*4.4.2. Translation Robustness Evaluation*

Table 12 (horizontal-shift) and Table 13 (vertical-shift) present the translation robustness results of the five models on the augmented test set consisting of 2500 samples. As depicted in Table 12, the horizontal robustness of all models has improved. Additionally, Table 13 indicates that vertical-robustness performances have also improved. This demonstrates that our data augmentation method effectively supplements the feature space, being beneficial for enhancing model robustness.

It is important to note that a vertical translation larger than 30 pixels could cause the top or bottom vertebra to cross image boundary, and such a scenario may introduce confusion for the models, as they could miss essential information for precise segmentation.

**Table 12.** DSC results (higher is better) of horizontal-shit robustness experiment of the five models. The displacement values are 10, 20, 30, 40 (pixels). An image could be shifted to the left and to the right. The results of left-shift and right-shift with the same displacement are averaged.

|  | 0 | 10 | 20 | 30 | 40 |
|---|---|---|---|---|---|
| SymTC | **96.169** ±1.146 | **96.169** ±1.152 | **96.168** ±1.157 | **96.167** ±1.165 | **96.165** ±1.173 |
| Swin UNETR | 94.115 ±7.661 | 94.088 ±7.744 | 94.014 ±8.018 | 93.904 ±8.429 | 93.725 ±9.049 |
| SLT-Net | 93.545 ±8.636 | 93.493 ±8.643 | 93.376 ±8.822 | 93.355 ±9.031 | 93.308 ±9.128 |
| UNETR | 94.642 ±4.260 | 94.639 ±4.300 | 94.617 ±4.439 | 94.461 ±5.404 | 94.052 ±7.824 |
| BianqueNet | 94.662 ±7.024 | 94.628 ±7.144 | 94.640 ±7.096 | 94.627 ±7.108 | 94.626 ±7.155 |

**Table 13.** DSC results (higher is better) of vertical-shit robustness experiment of the five models. The displacement values are 10, 20, 30, 40 (pixels). An image could be shifted up or down. The results of up-shift and down-shift with the same displacement are averaged.

|  | 0 | 10 | 20 | 30 | 40 |
|---|---|---|---|---|---|
| SymTC | **96.169** ±1.146 | **96.169** ±1.156 | **96.138** ±1.310 | **95.389** ±4.481 | 85.737 ±13.248 |
| Swin UNETR | 94.115 ±7.661 | 94.045 ±8.019 | 93.361 ±10.034 | 83.586 ±20.367 | 43.519 ±16.969 |
| SLT-Net | 93.545 ±8.636 | 93.235 ±9.180 | 92.334 ±10.872 | 86.659 ±17.703 | 68.053 ±22.927 |
| UNETR | 94.642 ±4.260 | 94.637 ±4.235 | 94.447 ±5.148 | 92.124 ±12.617 | 59.598 ±32.013 |
| BianqueNet | 94.662 ±7.024 | 95.165 ±5.578 | 95.299 ±5.161 | 94.430 ±7.430 | **92.777** ±10.775 |

*4.5. Ablation Study*

In this section, we perform various ablation experiments to investigate the significance and contribution of different configurations of hyperparameters, such as Transformer layers, attention heads, and embedding dimension, within our proposed model. These ablation experiments allow us to assess the effectiveness of our model in enforcing global structural regularity during the training phase, offering insights into the impact of different hyperparameter choices. In Tables 14-17, the bold settings indicate the default configuration in our SymTC. These variations of SymTC are trained on the original training set and tested on the augmented test set.

Table 14 presents the results of ablation study assessing the impact of Transformer path and CNN path within each SymTC module. The term "CNN only" refers to a configuration where all Transformer layers are turned off, and "Transformer only" indicates that all CNN layers are disabled. The results reveal that the configuration of the best instance segmentation performance is achieved by deactivating the Transformer path in TCM-2. It also reveals that the Transformer path is more crucial in the TC module when comparing the instance segmentation performance of "CNN only" and "Transformer only" configurations. This underlines the significant contribution of the Transformer path in enhancing the instance segmentation capabilities of SymTC. However, it is noted that the "CNN only" configuration exhibits superior translation robustness. Furthermore, turning off the CNN path in SymTC modules closer to the input would have a more negative impact on segmentation performance than doing so in modules near the output. There is no definitive evidence pointing to a specific TC module as being the most significant for improving either segmentation performance or translation robustness.

**Table 14.** DSC results (higher is better) of ablation study by switching on/off Transformer layers in SymTC. The other settings are unchanged. TCM-n represents the TC module #n in SymTC.

| SymTC Settings | Horizontal Translation | | | | Vertical Translation | | |
|---|---|---|---|---|---|---|---|
| | 0 | 10 | 20 | 40 | 10 | 20 | 40 |
| **CNN + Transformer** | 95.171 | 95.159 | 95.110 | **94.789** | 95.075 | 94.299 | 65.746 |
| CNN only | 81.366 | 81.564 | 81.477 | 81.551 | 82.000 | 81.946 | **81.530** |
| Transformer only | 93.166 | 93.146 | 93.066 | 92.389 | 93.098 | 92.059 | 45.748 |
| Disable Transformer in TCM-0 | 95.157 | 95.136 | 95.076 | 94.588 | 95.068 | 94.424 | 67.213 |
| Disable Transformer in TCM-1 | 95.086 | 95.082 | 95.008 | 94.689 | 95.063 | 93.841 | 48.990 |
| Disable Transformer in TCM-2 | **95.234** | **95.212** | 95.152 | 94.742 | **95.207** | 94.221 | 53.556 |
| Disable Transformer in TCM-3 | 95.040 | 95.026 | 94.948 | 94.409 | 95.041 | 93.762 | 49.539 |
| Disable CNN in TCM-0 | 94.093 | 94.098 | 94.069 | 93.883 | 94.036 | 93.4513 | 62.493 |
| Disable CNN in TCM-1 | 94.441 | 94.415 | 94.359 | 93.964 | 94.472 | 93.722 | 32.271 |
| Disable CNN in TCM-2 | 94.939 | 94.939 | 94.918 | 94.678 | 94.969 | 93.069 | 34.890 |
| Disable CNN in TCM-3 | 95.217 | 95.203 | **95.158** | 94.757 | 95.197 | **94.563** | 60.887 |

Table 15 summarizes the results obtained by varying the number of Transformer layers in SymTC modules for comparison purposes. The results show that the impact of the number of Transformer layers on instance segmentation performance is very minimal, indicating a few Transformer layers are good enough because a larger number of Transformer layers leads to higher computation cost. For translation robustness, it is evident that increasing the number of Transformer layers in each SymTC module leads to improved robustness in general.

**Table 15.** DSC results (higher is better) of ablation study by varying the number of Transformer layers in a TCM of SymTC. The other settings are unchanged.

| Num. of layers in a TCM | Horizontal Translation | | | | Vertical Translation | | |
|---|---|---|---|---|---|---|---|
| | 0 | 10 | 20 | 40 | 10 | 20 | 40 |
| **2** | 95.171 | 95.159 | 95.110 | 94.789 | 95.075 | 94.299 | 65.746 |
| 4 | 95.224 | 95.215 | 95.162 | 94.650 | 95.235 | 94.864 | 52.375 |
| 6 | 95.177 | 95.170 | 95.137 | 94.811 | 95.193 | **94.922** | 52.657 |
| 8 | **95.387** | **95.389** | 95.370 | 95.221 | **95.390** | 94.916 | 43.878 |
| 10 | 95.158 | 95.160 | 95.141 | 94.953 | 95.198 | 94.700 | 58.711 |
| 12 | 95.204 | 95.247 | 95.196 | 94.917 | 95.215 | 94.439 | 64.501 |
| 14 | 95.380 | 95.378 | 95.367 | 95.251 | 95.371 | 94.410 | 38.823 |
| 16 | 95.383 | 95.386 | **95.375** | **95.297** | 95.377 | 94.233 | 61.896 |
| 18 | 95.304 | 95.289 | 95.269 | 95.038 | 95.296 | 94.821 | **70.426** |

Table 16 presents the impact of head counts in the Multi-Head Self-Attention (MHSA) component of the SymTC modules. SymTC modules with a larger number of heads exhibit increased immunity to image shifting in the horizontal direction. The number of heads does not have much impact on the segmentation performance with zero-translation.

**Table 16.** DSC results (higher is better) of ablation study by varying the number of heads in Transformer layers of SymTC. The other settings are unchanged.

| Num. of heads | Horizontal Translation | | | | Vertical Translation | | |
|---|---|---|---|---|---|---|---|
| | 0 | 10 | 20 | 40 | 10 | 20 | 40 |
| 4 | 95.069 | 95.056 | 95.018 | 94.749 | 95.060 | 94.408 | 41.276 |
| 8 | 95.146 | 95.122 | 95.052 | 94.718 | 95.101 | 94.337 | 53.339 |
| **16** | 95.171 | 95.159 | 95.110 | 94.700 | 95.075 | 94.299 | 65.746 |
| 32 | 95.026 | 94.997 | 94.919 | 94.481 | 94.992 | 94.257 | 63.183 |
| 64 | **95.196** | **95.174** | 95.149 | 94.907 | **95.146** | 94.553 | **69.629** |
| 128 | 94.999 | 94.959 | 94.878 | 94.413 | 95.041 | 94.125 | 66.479 |
| 256 | 95.122 | 95.104 | **95.151** | **94.984** | 95.127 | **94.708** | 40.093 |

Table 17 shows the influence of the embedding dimension in Transformers within the SymTC modules. It demonstrates enhancement in segmentation performance and horizontal robustness with the increase in the embedding dimension. However, when the embedding dimension reaches 1024, there is a small drop in both segmentation performance and horizontal robustness, which suggests that increasing the embedding dimension beyond a certain threshold may lead to diminishing returns and negatively impact.

**Table 17.** DSC results (higher is better) of ablation study by varying the embedding dimension in Transformer layers of SymTC. The other settings are unchanged.

| Embedding Dimension | Horizontal Translation | | | | Vertical Translation | | |
|---|---|---|---|---|---|---|---|
| | 0 | 10 | 20 | 40 | 10 | 20 | 40 |
| 32 | 93.779 | 93.713 | 93.595 | 92.788 | 94.011 | 92.633 | **71.897** |
| 64 | 94.260 | 94.220 | 94.085 | 93.267 | 94.345 | 93.071 | 64.206 |
| 128 | 94.743 | 94.708 | 94.572 | 93.915 | 94.683 | 93.306 | 53.252 |
| 256 | 94.588 | 94.568 | 94.482 | 93.952 | 94.468 | 93.370 | 59.848 |
| **512** | **95.171** | **95.159** | **95.110** | **94.789** | 95.075 | 94.299 | 65.746 |
| 1024 | 95.104 | 95.080 | 95.007 | 94.625 | **95.091** | **94.429** | 57.398 |

## 5. Conclusion

In this paper, we present SymTC together with SSM-biomechanics based data augmentation, a dedicated approach to address the challenge of achieving precise instance segmentation of lumbar spine MR images. The performance of SymTC has been demonstrated through the comprehensive experiments and comparison with the other 15 existing models using our datasets, including the original dataset of 100 patients, and the generated SSMSpine dataset of thousands of virtual patients.

SymTC introduces a symbiotic relationship between CNN and Transformer that runs in two parallel paths in each TC module, which excels in assimilating both local and global contextual information from the spine MR images. This dual-path approach enables SymTC to achieve much better segmentation accuracy, compared to the models only using CNN or Transformer layers. The SymTC architecture also incorporates a novel relative position embedding that is specifically designed to fuse spatial information between content and position effectively, capturing and integrating critical spatial information. This relative position embedding elevates the model's capability in image segmentation.

We developed the SSM-biomechanics based data augmentation method to further improve model performance by providing large and diverse datasets of synthetic images with ground-truth labels. Given that our augmented datasets consist entirely of synthetic data, we have made our augmented dataset (SSMSpine) publicly available. The results presented indicate that models trained on the augmented training set had comparably or even better performance than the same models trained on the original training set. This underscores that our data augmentation method can generate synthetic data that eliminates privacy concerns while retaining in the same image domain.

Our current study mainly focused on the mid-sagittal lumbar spine MR images for two major reasons. First, as shown in a clinical study [83], the mid-sagittal image of a patient provides the most useful information for the diagnosis of lumbar spine degeneration. Secondly, the slice thickness of a lumbar MR scan in the sagittal direction is often much larger than 5mm, which causes difficulties to create accurate 3D ground-truth annotation for model training. Nevertheless, our model could be directly extended to handle 3D images once the sagittal slice thickness becomes acceptably small with the advancement of imaging technology.

## CrediT authorship contribution statement

**Jiasong Chen:** Data curation, Conceptualization, Formal analysis, Investigation, Methodology, Software, Validation, Visualization, Wring – original draft. **Linchen Qian:** Data curation, Investigation, Software, Writing – review & editing. **Linhai Ma:** Investigation, Writing – review & editing. **Timur Urakov:** Data curation, Writing – review & editing. **Weiyong Gu:** Writing – review & editing. **Liang Liang:** Methodology, Supervision, Project administration, Resources, Writing – review.

## Declaration of competing interest

The authors declare that they have no known competing financial interests or personal relationships that could have appeared to influence the work reported in this paper.